\documentclass[aoas,preprint]{imsart}

\usepackage{amsthm,amsmath,natbib}
\RequirePackage[colorlinks,citecolor=blue,urlcolor=blue]{hyperref}
\usepackage[toc,page]{appendix}
\usepackage[]{algorithmicx}
\usepackage{algorithm}
\usepackage[noend]{algpseudocode}
\usepackage{calc}
\usepackage{linegoal}
\usepackage{graphicx}
\usepackage{tikz}
\usepackage{pgfplots}
\pgfplotsset{compat=1.13}
\usepgfplotslibrary{fillbetween}
\usetikzlibrary{patterns}
\usepackage{amssymb}
\usepackage{rotating}
\usepackage{natbib,comment}
\usepackage{color,soul}
\usepackage{ulem}

\usepackage{xparse}
\usepackage{mathtools}
\usepackage{units}
\DeclarePairedDelimiter{\ceil}{\lceil}{\rceil}

\startlocaldefs

\def\X{\mathbf X}

\def\x{\mathbf x}

\def\y{\mathbf y}

\def\bbeta{\boldsymbol \beta}

\def\t{^\top}

\DeclareMathOperator*{\argmax}{argmax~} 
\DeclareMathOperator*{\argmin}{argmin~} 
\def \real{\rm I\!R}

\newcommand{\TB}{\textcolor{blue}}
\newcommand*{\Let}[2]{\State #1 $\gets$
\parbox[t]{\linegoal}{#2\strut}}
\newcommand*{\LongState}[1]{\State
\parbox[t]{\linegoal}{#1\strut}}
\endlocaldefs

\begin{document}

\begin{frontmatter}

\title{Qini-based Uplift Regression}
\runtitle{Qini-based Uplift Regression}

\begin{aug}
\author{\fnms{Mouloud} \snm{Belbahri,}\thanksref{t1}\ead[label=e1]{belbahrim@dms.umontreal.ca}}
\author{\fnms{Alejandro} \snm{Murua,}\corref{}\thanksref{t2}\ead[label=e2]{murua@dms.umontreal.ca}}
\author{\fnms{Olivier} 
\snm{Gandouet}\thanksref{t3}\ead[label=e3]{olivier.gandouet@tdassurance.com}}
\and
\author{\fnms{Vahid} 
\snm{Partovi Nia}\thanksref{t4}\ead[label=e4]{vahid.partovinia@polymtl.ca}}

\affiliation{Universit\'e de Montr\'eal\thanksmark{t1}\thanksmark{t2}, 
TD Insurance\thanksmark{t3} and \'Ecole Polytechnique de Montr\'eal\thanksmark{t4}}
\thankstext{t2}{Corresponding author: murua@dms.umontreal.ca}

\runauthor{Belbahri, Murua, Gandouet and Partovi Nia}

\address{D\'epartement de math\'ematiques et de statistique, \\
Universit\'e de Montr\'eal,\\
CP  6128, succ. centre-ville, \\
Montr\'eal, Qu\'ebec H3C 3J7 Canada\\
\printead{e1}}

\address{D\'epartement de math\'ematiques et de statistique, \\
Universit\'e de Montr\'eal,\\
CP  6128, succ. centre-ville, \\
Montr\'eal, Qu\'ebec H3C 3J7 Canada\\
\printead{e2}}

\address{Advanced Analytics, Research and Development,\\
  TD Insurance,\\
  50 Place Cr\'emazie,\\
  Montr\'eal, Qu\'ebec H2P 1B6 Canada\\
\printead{e3}}  

\address{Department of Mathematics and Industrial Engineering, \\
  \'Ecole Polytechnique de Montr\'eal, \\
2900  Boulevard \'Edouard Montpetit,\\ 
Montr\'eal, Qu\'ebec H3T 1J4 Canada\\
\printead{e4}}
\end{aug}

\begin{abstract}
Uplift models provide a solution to the problem of isolating the marketing effect of a campaign.
For customer churn reduction, uplift models are used to identify the customers who are likely to respond positively to a retention activity \textit{only} if targeted, and to avoid wasting resources on customers that are very likely to switch to another company.
We introduce a Qini-based uplift regression model to analyze a large insurance company's retention marketing campaign.
Our approach is based on logistic regression models.
We show that a Qini-optimized uplift model acts as a regularizing factor for uplift, much as
a penalized likelihood model does for regression.
This results in interpretable parsimonious models with few relevant explanatory variables.
Our results show that performing Qini-based parameters estimation significantly improves the uplift models performance.
\end{abstract}


\end{frontmatter}


\section{Introduction}\label{sec:data}

This work proposes a methodology that identifies characteristics associated with a home insurance policy that can be used to infer the
link between marketing intervention and policy renewal rate. Using the resulting statistical model, the goal is to predict which customers the company should focus on, in order to deploy future retention campaigns.

A subscription-based company loses its customers when they stop doing business with their service. Also known as customer attrition, customer churn can be a drag on the business growth. It is less expensive to retain existing customers than to acquire new customers, so businesses put effort into marketing strategies to reduce customer attrition. Customer loyalty, on the other hand, is usually more profitable because the company have already earned the trust and loyalty of existing customers. Businesses mostly have a defined strategy for mitigating customer churn. Organizations are able to determine their success rate in customer loyalty and identify improvement strategies using available data and learning about churn.

With the increasing amount of data available, a company tries to find the causal effects of customer churn. The term \textit{causal}, as in causal study, refers to a study that tries to discover a cause-effect relationship. The statement $A$ causes $B$ means that changing the value of $A$ will change the distribution of $B$. When $A$ causes $B$, $A$ and $B$ will be associated but the reverse is not, in general, true, since association does not necessarily imply causation. There exists two frameworks for discussing causation \citep{pearl2009causal}. We will consider the statistical framework for causal inference formally introduced by \cite{rubin1974estimating}, which uses the notation of counterfactual random variables. This framework is also associated with the potential outcome framework \citep{Neyman:1923}, also known as the Rubin causal model \citep{holland1986statistics}. Suppose a company decides to deploy a marketing campaign, and that customers are randomly divided into two groups. The first group is targeted with a marketing initiative (treatment group), and the second group serves as control (or baseline). A potential outcome is the theoretical response each customer would have manifested, had it been assigned to a particular group. Under randomization, association and causation coincide and these outcomes are independent of the assignment other customers receive. In practice, potential outcomes for an individual cannot be observed. Each customer is only assigned to either treatment or control, making direct observations in the other condition (called the counterfactual condition) and the observed individual treatment effects, impossible \citep{holland1986statistics}.

In marketing, the responses of customers in the treatment and control groups are observed. This makes it possible to calculate and compare the response rate of the two groups. A campaign is considered successful if it succeeds in increasing the response rate of the treated group relative to the response rate of the control group. The difference in response rate is the increase due to the campaign. To further increase the returns of future direct marketing campaigns, a predictive response model can be developed. Response models \citep{smith1982information, hanssens2003market, coussement2015improving} of client behavior are used to predict the probability that a client responds to a marketing campaign (e.g. renews subscription). Marketing campaigns using response models concentrate on clients with high probability of positive response. However, this strategy does not necessarily cause the renewal. In other words, the customers could renew their subscription without marketing effort. Therefore, it is important to extract the cause of the renewal, and isolate the effect of marketing.

Data from one of the leader north-American insurers is at our disposal to evaluate the performance of the methodology introduced in this work. This company is interested in designing retention strategies to minimize its policyholders' attrition rate. For that purpose, during three months, an experimental loyalty campaign was implemented, from which policies coming up for renewal were randomly allocated into one of the following two groups: a treatment group, and a control group. Policyholders under the treatment group received an outbound courtesy call made by one of the company's licensed insurance advisors, with the objective to reinforce the customers confidence in the company, to review their coverage and address any questions they might have about their policy. No retention efforts were applied to the control group. The goal of the study is develop
models that will be used to identify which clients are likely to benefit from a call at renewal, that is, clients that are likely to renew their policy \textit{only} if they are called by an advisor during their renewal period. Also, clients that are not targeted will most likely
\begin{itemize}
    \item renew their policy on their own,
    \item cancel their policy whether they receive a call or not,
    \item cancel their policy \textit{only} if they receive a call.
\end{itemize}
Table~\ref{tab:randomization} shows the marketing campaign retention results. The observed difference in retention rates between the treated group and the control group is small, but there is some evidence of a slightly negative impact of the outbound call. Even if the difference is slightly negative, it may be the case that the campaign had positive retention effects on some subgroup of customers, but they were offset by negative effects on other subgroups. This can be explained by the fact that some customers are already dissatisfied with their insurance policies and have already decided to change them before receiving the call. This call can also trigger a behavior that encourages customers to look for better rates.
\begin{table}[H]
    \centering
    \caption{Renewal rate by group for $n=20,997$ home insurance policies.}
    \begin{tabular}{lccc}
     \hline
                       & Control     &  Called      & Overall \\
     \hline
    Renewed policies   & $2,253$     &  $18,018$    & $20,271$ \\
    Cancelled policies & $72$        &  $654$       & $726$\\
    Renewal rate       &  $96.90\%$  &  $96.50\%$   & $96.54\%$ \\
    \hline
    \end{tabular}
    \label{tab:randomization}
\end{table}

In a randomized experiment, researchers often focus on the estimation of average treatment effects; the effect of the marketing initiative on a particular client is determined from this estimate. However, there might be a proportion of the customers that responds favorably to the marketing campaign, and another proportion that does not. A decision based on an average treatment effect at the individual level would require an adjustment because of the heterogeneity in responses that can be originated by many factors.

The so-called uplift model \citep{radcliffe1999differential, hand2001idiot, lo2002true} provides a solution to the problem of isolating the marketing effect. Instead of modeling the class probabilities, uplift attempts to model the difference between conditional class probabilities in the treatment (e.g., a marketing campaign) and control groups. Uplift modeling aims at identifying groups on which a predetermined action will have the most positive effect.

Assessing model performance is complex for uplift modeling, as the actual value of the response, that is, the \textit{true} uplift, is unknown at the individual subject level. To overcome this limitation, one can assess model performance by comparing groups of observations.
This is done through the Qini coefficient \citep{radcliffe2007using}, which plays a similar role as the
Gini coefficient \citep{gini1997concentration} in Economics.
The Qini coefficient is a single statistics drawn form the Qini curve. This latter object 
is a generalization of the Lorenz curve \citep{lorenz1905methods}
traditionally used in direct marketing for response models.

As in all regression-based modeling, an issue in uplift modeling is the ease of interpretation of the results. The model becomes harder to interpret when the number of potential explanatory variables, that is the \textit{dimension} of the explanatory variables increases. When the variable dimension is small, knowledge-based approaches to select the optimal set of variables can be effectively applied. When the number of potentially important variables is too large, it becomes too time-consuming to apply a manual variable selection process. In this case one may consider using automatic subset selection tools. Variable selection is an important step. It reduces the dimension of the model, avoids overfitting, and improves model stability and accuracy \citep{guyon2003introduction}. Well-known variable selection techniques such as forward, backward, stepwise \citep{montgomery2012introduction}, stagewise \citep{hastie2007forward}, lasso \citep{Tibshirani_lasso_1996}, and LARS \citep{efron2004least}, among others, are not designed for uplift models. One might need to adapt them to perform variable selection in this context.

We propose a new way to perform model selection in uplift regression models.
Our methodology is based on the maximization of a modified version of the Qini coefficient, {\it the adjusted Qini}, that we introduced in Section~\ref{sec:adjusted_qini}.
Because model selection corresponds to variable selection, the task is haunting and intractable if done in a straightforward manner when the number of variables to consider is large, e.g. $p\approx 100$, like in the case of the insurance data.
To realistically search for a good model, we conceived a searching method based on an efficient exploration of the regression coefficients space combined with a lasso penalization of the log-likelihood. There is no explicit analytical expression for the adjusted Qini surface (nor for the Qini curve), so unveiling it is not easy. Our idea is to gradually uncover the adjusted Qini surface in a manner inspired by surface response designs. The goal is to find the global maximum or a reasonable local maximum of the adjusted Qini by exploring the surface near optimal values of the coefficients. These coefficient values are given by maximizing the lasso penalized log-likelihood. The exploration is done using Latin hypercube sampling structures \citep{mckay2000comparison} centered in a sequence of penalized estimates of the coefficients.

The rest of the paper is organized as follows. We first present the current uplift models in Section \ref{sec:literature} and Section \ref{sec:metho} introduces the notation and details of Qini-based uplift regression. Sections \ref{sec:simulation} and \ref{sec:analysis} present the computational results of the proposed methodology on synthetic and real datasets. Final remarks and conclusion are given in Section \ref{sec:conclusion}.

\section{Uplift modeling}\label{sec:literature}

Let $Y$ be the 0-1 binary response variable, $T$ the 0-1 treatment indicator variable and $X_1,\ldots,X_p$ the explanatory variables (predictors). The binary variable $T$ indicates if a unit is exposed to treatment ($T=1$) or control ($T=0$). Suppose that $n$ independent units are observed $\{(y_i, \x_i, t_i)\}_{i=1}^n$, where $\x_i = (x_{i1}, \ldots, x_{ip})$ are realisations of the predictors random variables.
For $i=1,\ldots,n$, an uplift model estimates
\begin{equation}
u (\textbf{x}_i)=\mathrm{Pr}(Y_i = 1 \mid \x_i ,T_i=1) - \mathrm{Pr}(Y_i =1 \mid \x_i  ,T_i=0), 
\end{equation}
where the notation $\mathrm{Pr}(Y_i = y_i \mid \x_i ,T_i=t_i)$ stands for the corresponding conditional probability.
Uplift modeling was formally introduced in \cite{radcliffe1999differential} under the appellation of {\it differential response modeling} where a thorough motivation and several practical cases promoted uplift modeling in comparison with common regression or basic tree-based methods that were used to predict the probability of success for the treatment group. They showed that conventional models, which were referred to as {\it response models}, did not target the people who were the most positively influenced by the treatment. In \citep{radcliffe1999differential} and \citep{hansotia2002incremental}, the methods introduced are tree-based algorithms similar to CART \citep{breiman1984classification}, but using modified split criteria that suited the uplift purpose. The method proposed by \cite{hansotia2002incremental} uses the uplift's absolute difference $\Delta = |u_l - u_r|$, where $u_l, u_r$ are the observed uplifts in the left and right child nodes, respectively. It is also possible to use the difference in node sizes as some sort of penalty term to adjust the differences in uplift \citep{radcliffe2011real}. Other split criteria proposed in the literature are based on the $\chi^2$ statistic \citep{su2009subgroup,radcliffe2011real}, which is usually a function of $\Delta^2$. All these splitting criteria rely on maximizing heterogeneity in treatment effects ($\Delta$).

\subsection{Adjusted Qini}
\label{sec:adjusted_qini}

Evaluating uplift models requires the construction of the {\it Qini curve} and the computation of the {\it Qini coefficient} \citep{radcliffe2007using}.
The motivation to consider the Qini curve comes from the fact that
a good model should be able to select individuals with highest uplift first.
More explicitly, for a given model, let $\hat{u}_{(1)} \geq \hat{u}_{(2)} \geq ... \geq \hat{u}_{(n)}$ be the sorted predicted uplifts.
Let $\phi \in [0,1]$ be a given proportion and let
$ N_{\phi} = \{i: \hat{u}_{i} \geq \hat{u}_{(\ceil{\phi n})} \} \subset \lbrace 1, \ldots, n \rbrace$
be the subset of individuals with the $\phi n \times 100 \%$ highest predicted uplifts $\hat{u}_i$
(here $\ceil{s}$ denotes the smallest integer larger or equal to $s\in \real$).
Because $N_{\phi}$ is a function of the predicted uplifts, $N_{\phi}$ is a function of the fitted model. For a parametric model such as (\ref{eq:uplift_likelihood}), $N_{\phi}$ is a function of the model's parameters estimates, and should be denoted $N_{\phi} (\hat{\boldsymbol{\theta}})$. To simplify the notation, we prefer to omit this specification.

As a function of the fraction of population targeted $\phi$, the incremental uplift is defined as
$$ h(\phi) = \sum\limits_{i \in  N_{\phi}} y_i t_i - \sum\limits_{i \in  N_{\phi}} y_i (1-t_i) \biggl\{ \sum\limits_{i \in  N_{\phi}} t_i / \sum\limits_{i \in  N_{\phi}} (1-t_i) \biggr\}, $$
where $\sum_{i \in  N_{\phi}} (1-t_i) \neq 0$, with $h(0)=0$. The incremental uplift has been normalized by the number of subjects treated in $N_\phi$.
The relative incremental uplift $g(\phi)$ is given by $g(\phi) = h(\phi) / \sum_{i=1}^n t_i.$ Note that $g(0)=0$ and $g(1)$ is the overall sample observed uplift $$ g(1) = \biggl(\sum\limits_{i=1}^n y_i t_i / \sum\limits_{i=1}^n t_i \biggr) - \biggl( \sum\limits_{i=1}^n y_i (1-t_i) / \sum\limits_{i=1}^n (1-t_i) \biggr).$$

\begin{figure}
    \centering
    \begin{tikzpicture}
        \begin{axis}[ylabel=$g(\phi)$: Relative Incremental Uplift ,xlabel=$\phi$: Proportion of Population Targeted ,
        xmin=0, xmax=1,ymin=-1.2, ymax=0.7, samples=500]
            \addplot[name path=F,dashed, ultra thick,domain={0:1}] {-2*x*x + x};
            \addplot[name path=G,dotted, ultra thick,domain={0:1}] {-3*x*x + 2*x};
            \addplot[name path=K,gray,domain={0:1}] (x, -x); 
            \addplot[pattern=north west lines, pattern color=brown!50]fill between[of=F and K, soft clip={domain=0:1}];
            \addplot[pattern=north west lines, pattern color=blue!10]fill between[of=G and K, soft clip={domain=0:1}];
        \end{axis}
        
        \draw [-, dotted, ultra thick] (4.1,5.5)--(4.5,5.5) node [right] {\small Model 1};
        \draw [-, dashed, ultra thick] (4.1,5.2)--(4.5,5.2) node [right] {\small Model 2};
        \draw [-, gray] (4.1,4.9)--(4.5,4.9) node [right] {\small Random};
    \end{tikzpicture}   
    \caption{Example of Qini curves corresponding to two different uplift models compared to a random targeting strategy.}
    \label{fig:qinicurve}
\end{figure}
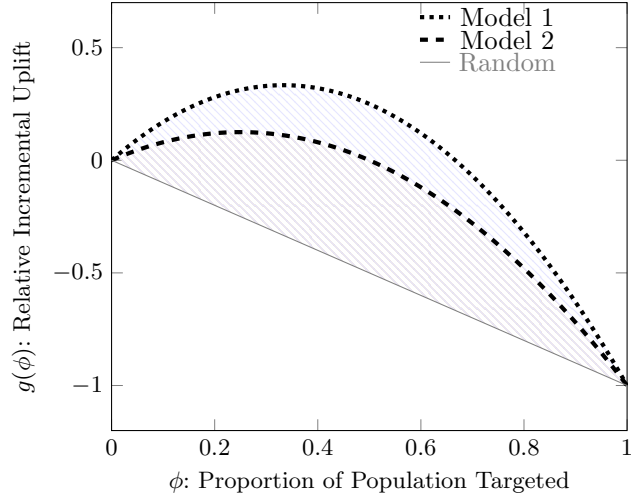

The Qini curve is constructed by plotting $g(\phi)$ as a function of $\phi \in [0,1]$. This is illustrated in Figure \ref{fig:qinicurve}.
The curve can be interpreted as follows. The $x$-axis represents the fraction of targeted individuals and the $y$-axis shows the incremental number of positive responses relative to the total number of targeted individuals. The straight line between the points $(0,0)$ and $(1, g(1))$ in Figure~\ref{fig:qinicurve} represents a benchmark to compare the performance of the model to a strategy that would randomly target subjects. In other words, when the strategy is to treat individuals randomly, if a proportion $\phi$ of the population is treated, we expect to observe an uplift equal to $\phi$ times the global uplift. The Qini coefficient $q$ is a single index of model performance. It is defined as the area between the Qini curve and the straight line
\begin{equation}
     q = \int_0^1 Q(\phi) ~\mathrm{d}\phi = \int_0^1 \{g(\phi) - \phi~g(1)\} ~\mathrm{d}\phi,
     \label{eq:q_coeff}
\end{equation}
where $Q(\phi) = g(\phi) - \phi~g(1)$. This area can be numerically approximated using a Riemann method such as the trapezoid rule formula: the domain of $\phi \in [0,1]$ is partitioned into $J$ panels, or $J+1$ grid points $0=\phi_1 < \phi_2 < ... < \phi_{J+1} = 1$, to approximate the Qini coefficient $q$ (\ref{eq:q_coeff}) by its empirical estimation
\begin{equation}
  \hat{q} = \dfrac{1}{2} \sum_{j=1}^J (\phi_{j+1}-\phi_j)\{Q(\phi_{j+1}) + Q(\phi_{j})\}.
  \label{eq:q:hat}
\end{equation}
In general, when comparing several models, the preferred model is the one
with the maximum Qini coefficient \citep{radcliffe2007using}.

\begin{figure}
    \centering
    \begin{tikzpicture}[scale=0.75]
    \begin{axis}[
            ybar,
            bar width=1cm,
            symbolic x coords={0--20,20--40,40--60,60--80,80--100},
            xtick=data,
            xlabel={Population Targeted (\%)},
            ylabel={Uplift (\%)},
        ]
        \addplot[red!20!black,fill=black!20!white] table[x=interval,y=uplift]{\databarplot};
    \end{axis}
    \end{tikzpicture}
    \hspace{0.1cm}
    \begin{tikzpicture}[scale=0.75]
     \begin{axis}[
            ybar,
            bar width=1cm,
            symbolic x coords={0--20,20--40,40--60,60--80,80--100},
            xtick=data,
            xlabel={Population Targeted (\%)},
            ylabel={Uplift (\%)},
        ]
        \addplot[red!20!black,fill=black!20!white] table[x=interval,y=uplift2]{\databarplot};
    \end{axis}
    \end{tikzpicture}
    \caption{Theoretical predicted uplift barplots with $5$ panels corresponding to two different models. A good model should order the observed uplift from highest to lowest. The Kendall's uplift rank correlation is $\rho=1$ for the left barplot and $\rho=0.4$ for the right barplot.}
    \label{fig:qinibarplot}
\end{figure}
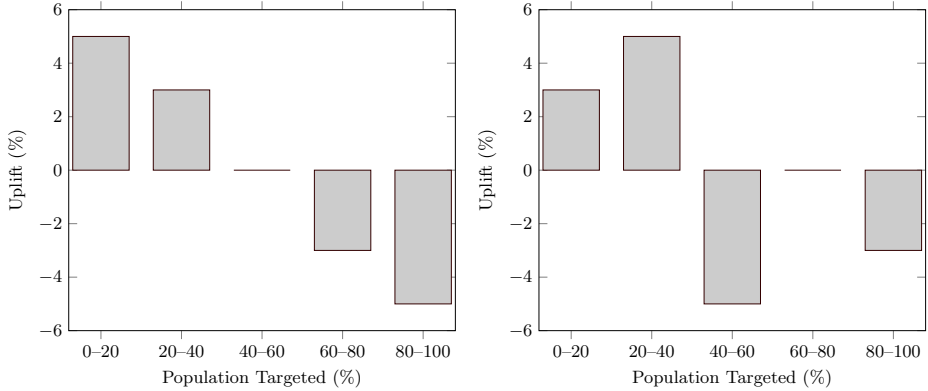

Another visualization associated with uplift model validation is based on the observed uplifts in each of the $J$ bins used to compute the Qini coefficient: a good model should induce a decreasing disposition of the observed uplifts in these bins. Figure \ref{fig:qinibarplot} illustrates good and bad uplift models as barplots of observed uplifts associated with each of the $J$ bins. A decreasing disposition of the uplift values in the $J$ bins is an important property of an uplift model. To measure the degree to which a model does this correctly, we suggest the use of the Kendall rank correlation coefficient \citep{kendall1938new}. The goal is to find a model that maximizes the correlation between the predicted uplift and the observed uplift. The Kendall's uplift rank correlation is defined as
\begin{equation}
    \rho = \frac{2}{J(J-1)} \sum_{i<j} \mathrm{sign}(\bar{\hat{u}}_i - \bar{\hat{u}}_j)~\mathrm{sign}(\bar{u}_i - \bar{u}_j),
    \label{eq:corr_coeff}
\end{equation}
where $\bar{\hat{u}}_k$ is the average predicted uplift in bin $k$,  $k \in {1,...,J}$, and $\bar{u}_k$ is the observed uplift in the same bin.

\pgfplotstableread[row sep=\\,col sep=&]{
    interval & uplift & uplift2 \\
    0--20     & 5 & 3 \\
    20--40     & 3 & 5\\
    40--60    & 0 & -5\\
    60--80   & -3  & 0\\
    80--100   & -5 & -3 \\
    }\databarplot

\begin{figure}
    \centering
    \begin{tikzpicture}[scale=0.75]
    \begin{axis}[
            ybar,
            bar width=1cm,
            symbolic x coords={0--20,20--40,40--60,60--80,80--100},
            xtick=data,
            xlabel={Population Targeted (\%)},
            ylabel={Uplift (\%)},
        ]
        \addplot[red!20!black,fill=black!20!white] table[x=interval,y=uplift]{\databarplott};
    \end{axis}
    \end{tikzpicture}
    \hspace{0.1cm}
    \begin{tikzpicture}[scale=0.75]
     \begin{axis}[
            ybar,
            bar width=1cm,
            symbolic x coords={0--20,20--40,40--60,60--80,80--100},
            xtick=data,
            xlabel={Population Targeted (\%)},
            ylabel={Uplift (\%)},
        ]
        \addplot[red!20!black,fill=black!20!white] table[x=interval,y=uplift2]{\databarplott};
    \end{axis}
    \end{tikzpicture}
    \caption{Theoretical predicted uplift barplots with $5$ panels corresponding to two different models. The left panel model has a much smaller value of $\hat{q}$ than the one on the right panel. However, $\rho=1$ for the left panel and $\rho=0.8$ for the right panel.}
    \label{fig:kendall_example}
\end{figure}
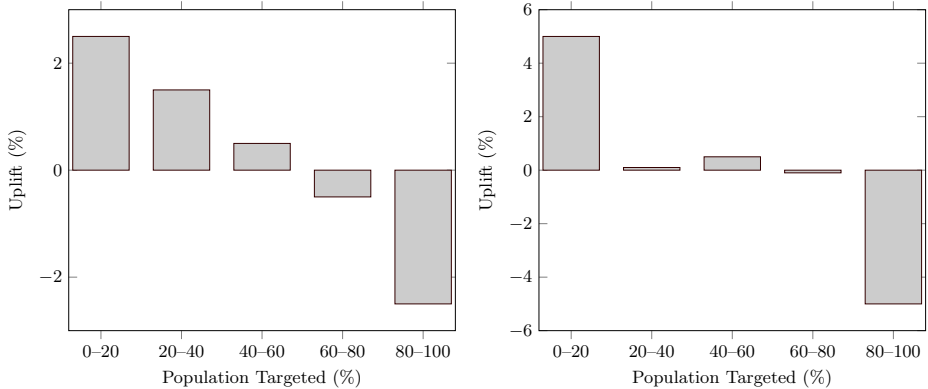

From a business point of view, this statistic and the associated barplot are easier to interpret
than the Qini and the Qini curve.
However, we do not advise the use of  $\rho$ alone for model selection.
If two models have the same $\hat q$, the preferred one should be the one with the highest $\rho$.
But, when two models $\hat q$ differ, it is not clear that the preferred one should be the one with the highest $\rho$.
In Figure~\ref{fig:kendall_example}, we show an example of two models where the one in the left has a perfect Kendall's uplift
rank correlation ($\rho=1$) but with a Qini coefficient much smaller than the model on the right panel.
In this scenario, the model with $\rho=0.8$ is the best.

We propose an appropriate combination of (\ref{eq:q:hat}) and (\ref{eq:corr_coeff}): the \textit{adjusted Qini coefficient} which is
given by
\begin{equation}
    \hat{q}_{\mathrm{adj}} = \rho~\mathrm{max}\{ 0, \hat{q}\}.
    \label{eq:qadj}
\end{equation}
The adjusted Qini coefficient represents a trade-off between maximizing the area under the Qini curve and grouping the individuals in decreasing uplift bins.

%
\pgfplotstableread[row sep=\\,col sep=&]{
    interval & uplift & uplift2 \\
    0--20     & 2.5 & 5 \\
    20--40     & 1.5 & 0.1\\
    40--60    & 0.5 & 0.5\\
    60--80   & -0.5  & -0.1\\
    80--100   & -2.5 & -5 \\
    }\databarplott

\paragraph{A note on the estimation of the Qini curve}
The number of bins $J$ may be seen as a hyper-parameter. Its choice will certainly affect the computation of the adjusted Qini coefficient. In practice, the sample is divided into quantiles ($J=5$) or deciles ($J=10$).
In order to have a hint on what adequate values for $J$ are, 
suppose that the relative incremental uplift function $g(\phi)$ is twice-differentiable, with bounded second derivative.
Consider the trapezoid rule approximation
to the integral $q$ based on $J$ bins. Let us assume that the bin sizes are proportional to $1/J$.
It is well-known that under these assumptions the error of the approximation is order ${\cal O}(1/J^2)$.
Since $g(\cdot)$ is unknown, one need to estimate it with data. Let $\hat g_j$ be the estimate of $g( \phi_j)$, $j=1,\ldots, J.$ Suppose that $\hat g_j$ is obtained as a mean of $n/J$ random variables observed in the $j$-th bin. We suppose that these random variables are independent and identically distributed
with mean $g( \phi_j)$ and a certain finite variance. The weak law of large numbers says that $\hat g_j$ converges to $g( \phi_j)$, and the error in this approximation is of order ${\cal O}(J/\sqrt{n})$. It turns out that we need $J$ to minimize   $\kappa_1/J^2 + \kappa_2 J/\sqrt{n}$, where $\kappa_1, \kappa_2$ are constants. The solution is $J = {\cal O}(n^{1/6})$. So, for example, if $n \approx 1000$, then the optimal $J \approx 3$. Hence, the usual values of $J=5$ and $J=10$ seem reasonable to estimate the Qini \citep{radcliffe2007using}.

\subsection{Brief overview of previous work on uplift modeling}\label{sec:literature2}

The intuitive approach to uplift modeling is to build two separated classification models. \cite{hansotia2002direct} used the {\it two-model} approach which consists in direct subtraction of models for the treated and untreated groups.
The asset of this technique is its simplicity. However, in many cases this approach performs poorly \citep{radcliffe2011real}.
Both models focus on predicting the class probabilities instead of making the best effort to predict the uplift, i.e., the difference between two probabilities. General discussions following differential response modeling and the two-model approach appeared in \cite{hansotia2002incremental} where the technique known as {\it incremental value modeling} was introduced. This uses the difference in response rates in the two groups (treatment and control) as the split criterion of a regression tree. Also, \cite{lo2002true} introduced the {\it true lift modeling} using a single standard logistic regression model which explicitly added interaction terms between each explanatory variable and the treatment indicator. The interaction terms measure the additional effect of each explanatory variable because of treatment. The model yields an indirect estimation of the causal effect by subtracting the corresponding prediction probabilities, which are obtained by respectively
setting the treatment indicator variable to treated and control in the fitted model.
The disadvantage with this solution is that it is not optimized with respect to the goodness-of-fit measures designed for uplift. Instead, the parameters are estimated with respect to the likelihood. Our results show that estimating the regression parameters by maximizing the adjusted Qini significantly improves the uplift models performance.

Most current approaches that directly model the uplift causal effect are adaptations of classification and regression trees \citep{breiman1984classification}. \cite{rzepakowski2010decision} propose a tree-based method based on generalizing classical tree-building split criteria and pruning methods. The approach is based on the idea of comparing the distributions of outcomes in treatment and control groups, using a divergence statistic,
such as the Kullback-Leibler divergence or a modified Euclidean distance \citep{rzepakowski2012decision,guelman2012random,rzepakowski2010decision}.
Another non-parametric method is discussed in \citep{Alemi.etal-PersonalizedMedicine-2009,su2012facilitating}. Therein the uplift is estimated from the nearest neighbors containing at least one treated and one control observation. This method quickly becomes computationally expensive when dealing with large datasets, because the entire dataset has to be stored in order to predict the uplift for new observations. For a more detailed overview of the uplift modeling literature, the reader is referred to the works of \cite{kane2014mining}, \cite{gutierrez2017causal} and \cite{devriendt2018literature}.

From a complexity point of view, parametric models are simpler than non-parametric ones such as regression trees, because  for parametric models, the number of parameters is kept small and fixed.
Although, for many analysts prediction is the main target, from a business point of view, model interpretation is very important.
Knowing which variables and how these variables discriminate between groups of clients is one of the main goal of uplift modeling for marketing.
For these reasons, in this work we focus on parametric models.
We develop our methodology for the logistic regression since interpretation of the odds ratios is well-known.
However, our estimation procedure can be easily generalized to other parametric models.

\section{Qini-based logistic regression for uplift}\label{sec:metho}

Logistic regression is a well-known parametric model for binary response variables. Given a $p$-dimensional predictor vector $\textbf{x}_i$, $i \in \{1,\ldots,n\},$ logistic intercept $\theta_o \in \real$, and logistic regression coefficients $\bbeta \in \real^p$, 
the model is
\begin{equation*} 
     p_i = p_i( \theta_o, \bbeta) = \mathrm{Pr}(Y_i=1 \mid \textbf{x}_i, \theta_o, \bbeta) = \bigl(1 + \mathrm{exp}\{- (\theta_o + \textbf{x}_i^{\top} \bbeta) \} \bigr)^{-1}
\end{equation*}
or, equivalently,
$\mathrm{logit}(p_i) = \theta_o + \textbf{x}_i^{\top} \bbeta,$
where 
$\mathrm{logit}(p_i) = \log\{ p_i / (1-p_i)\}.$
Throughout the paper, the superscript $^\top$ stands for the transpose of a column vector or matrix.
In the uplift context, one need to add explicit interaction terms between each explanatory variable and the treatment indicator.
Let $\gamma$ denote the treatment effect, $\boldsymbol{\beta}$, the vector of main effects, $\boldsymbol{\delta}$, the vector of interactions effects, and $\theta_o$, the intercept. 
The model is 
\begin{equation}
  p_i(\theta_o, \boldsymbol{\theta}) = \mathrm{Pr} (Y_i = 1 \mid \mathbf{x}_i, t_i, \theta_o, \boldsymbol{\theta}) =
  \bigl( 1+ \mathrm{exp}\{-(\theta_o + \gamma t_i + \mathbf{x}_i^\top [\boldsymbol{\beta} + t_i \boldsymbol{\delta}] ) \} \bigr)^{-1},
\label{eq:pitheta}
\end{equation}
where $\boldsymbol{\theta} = (\boldsymbol{\beta}, \gamma, \boldsymbol{\delta})$, denotes all model parameters except for the intercept $\theta_o$.
The likelihood function associated with the uplift model is 
\begin{equation}
    \mathcal{L}(\theta_o, \boldsymbol{\theta}) =  \prod_{i=1}^{n} p_i(\theta_o, \boldsymbol{\theta})^{y_i} \{1-p_i(\theta_o, \boldsymbol{\theta})\}^{(1-y_i)},
    \label{eq:uplift_likelihood}
\end{equation}
where $\{y_i : i=1,\ldots,n\}$ are the observed response variables.
The maximum likelihood estimates of  $(\theta_o, \boldsymbol{\theta})$ will be denoted by $(\hat{\theta}_o, \hat{\boldsymbol{\theta}}),$
with $\hat{\boldsymbol{\theta}} = (\boldsymbol{\hat{\beta}}, \hat{\gamma}, \boldsymbol{\hat{\delta}})$.
The predicted uplift associated with the covariates vector $\mathbf{x}_{n+1}$ of a future individual is estimated by
$$ \hat{u}(\mathbf{x}_{n+1}) = \bigl(1+\mathrm{exp}\{-(\hat{\theta}_o + \hat{\gamma} + \mathbf{x}_{n+1}^\top[ \boldsymbol{\hat{\beta}} + \boldsymbol{\hat{\delta}}] )\}\bigr)^{-1} - \bigl(1+ \mathrm{exp}\{-(\hat{\theta}_o + \mathbf{x}_{n+1}^\top \boldsymbol{\hat{\beta}})\}\bigr)^{-1}.$$

We propose to select a regression model that maximizes the adjusted Qini coefficient. To realistically search for a good model, we conceived a searching method based on
Latin hypercube sampling of the regression coefficients space combined with a lasso penalization of the log-likelihood. The procedure is explained in the following sections.

\subsection{Estimation of the Qini maximizer}

Because the adjusted Qini function is not straightforward to optimize with respect to the parameters, one needs to explore the parameters space in order to find the maximum of the adjusted Qini.

Latin hypercube sampling (LHS) is a statistical method for quasi-random sampling based on a multivariate probability law inspired by the Monte Carlo method \citep{mckay2000comparison}. The method performs the sampling by ensuring that each sample is positioned in a space $\Omega$ of dimension $p$ as the only sample in each hyperplane of dimension $p-1$ aligned with the coordinates that define its position. Each sample is therefore positioned according to the position of previously positioned samples to ensure that they do not have any common coordinates in the $\Omega$ space.
When sampling a function of $p$ variables, the range of each variable is divided into $M$ equally probable intervals. $M$ sample points are then placed to satisfy the Latin hypercube requirements; this forces the number of divisions, $M$, to be equal for each variable. Also this sampling scheme does not require more samples for more dimensions (variables); this independence is one of the main advantages of this sampling scheme.
We use LHS to find the coefficient parameters that maximize the adjusted Qini. The procedure to search for the Qini maximizer is explained next.
It is based on the lasso penalized likelihood and several LHS structures.

\subsubsection{Penalized log-likelihood}
In the context of linear regression, the effectiveness of penalization has been amply supported practically and theoretically in several studies. In order to decrease the mean squared error of least squares estimates, ridge regression \citep{hoerl1970ridge} has been proposed as a trade-off between bias and variance. This technique adds an $L_2$-norm penalization term to the least squares loss. The {\it lasso} (least absolute shrinkage and selection operator) penalization technique \citep{Tibshirani_lasso_1996} uses an $L_1$-norm penalization which sets some of the regression coefficients to zero (sparse selection) while shrinking the rest. The elastic net penalization technique \citep{zou2005regularization} linearly combines the $L_1$ and $L_2$-norms to provide better prediction in the presence of collinearity. Other penalization techniques such as {\it scad} \citep{fan2001variable} and bridge regression \citep{frank1993statistical}, offer interesting theoretical properties, including consistency. 

Here, we focus on sparse estimation of the coefficients. That is, the selection of a small subset of features to predict the response. This is often achieved with a $L_1$-norm penalization.
Given $\lambda \in \real^+$, in the context of linear regression,
the {\it lasso} penalization \citep{Tibshirani_lasso_1996} finds the estimate of the coefficients
$\hat{\bbeta}(\lambda)$ that maximizes the penalized log-likelihood, say $\ell(\beta) + \lambda \sum_{j=1}^p |\beta_j|.$
Setting the penalization constant $\lambda=0$ returns the least squares estimates which performs no shrinking and no selection.
For
$\lambda>0$, the regression coefficients $\hat\bbeta(\lambda)$ are shrunk towards zero, and some of them are set to zero (sparse selection).
\cite{friedman2007pathwise} proposed a fast pathwise coordinate descent method to find $\hat\bbeta(\lambda)$, using the current estimates as warm starts. In practice, the value of $\lambda$ is unknown. Cross-validation is often used to search for a good value of the penalization constant. 
The least angle regression (or LARS algorithm) efficiently computes a path of values of $\hat\bbeta(\lambda)$
over a sequence of values of $\lambda=\lambda_1< \cdots < \lambda_j < \cdots < \lambda_{\min(n,p)},$ for which the parameter dimension changes \citep{Efronetal_lar_2004}. The entire sequence of steps in the LARS algorithm
with $p < n$ variables requires $\mathcal{O}(p^3 + np^2)$ computations, which is the cost of a single least squares fit on $p$ variables.
Extensions to generalized linear models with nonlinear loss functions require some form of approximation. In particular, for the logistic regression case, which is our model of interest, \cite{friedman2010regularization} extend the pathwise coordinate descent algorithm \citep{friedman2007pathwise} by first, approximating the log-likelihood (quadratic Taylor expansion about current estimates), and then using coordinate descent to solve the penalized weighted least-squares problem. The algorithm computes the path of solutions for a decreasing sequence of values for $\lambda=\lambda_{\min(n,p)} > \cdots > \lambda_j > \cdots > \lambda_1,$ starting at the smallest value for which the entire vector $\hat\bbeta = 0$. The algorithm works on large datasets, and is publicly available through the \textbf{R} package \textit{glmnet} \citep{friedman2009glmnet}, which we use in this work. In what follows, we will refer to the sequence of regularizing constant values given by \textit{glmnet} as the \textit{logistic-lasso sequence}.

\subsubsection{Qini-optimized uplift regression}
\label{sec:qini:pur}

Recall the uplift model likelihood given in (\ref{eq:uplift_likelihood}).
The vector of parameters $\boldsymbol{\theta} = (\bbeta, \gamma, \boldsymbol{\delta})$
is a $p'=(2p+1)$-dimensional vector.
Because of the considerations mentioned in the previous sections, 
in order to select an appropriate
sparse model for uplift, we adapt the lasso
algorithm to explore a relatively small set of reasonable models, so as to avoid an
exhaustive model search.
The penalized uplift model log-likelihood is given by  
\begin{equation}
    \ell (\theta_o, \boldsymbol{\theta} \mid \lambda) =  \sum_{i=1}^{n} \big( y_i\mathrm{log} \{p_i(\theta_o, \boldsymbol{\theta})\} + (1-y_i) \mathrm{log} \{1-p_i(\theta_o, \boldsymbol{\theta})\} \big) + \lambda \Vert \boldsymbol{\theta} \Vert_1,
    \label{eq:uplift_loglikelihood}
\end{equation}
where $p_i(\theta_o, \boldsymbol{\theta})$ is as in (\ref{eq:pitheta}), and $\Vert \cdot \Vert_1$ stands for the $L_1$-norm.
For any given $\lambda$, the parameters that maximize the penalized log-likelihood (\ref{eq:uplift_loglikelihood}) are denoted by
\begin{equation}
    \bigl(\hat{\theta}_o(\lambda), \hat{\boldsymbol{\theta}}(\lambda) \bigr) = \argmax_{\theta_o, \boldsymbol{\theta}} \ell (\theta_o, \boldsymbol{\theta}\mid \lambda).
    \label{eq:maxloglikelihood}
\end{equation}
Applying the pathwise coordinate descent algorithm to the uplift model, we get a sequence of critical penalization values
$\lambda_1 < \cdots < \lambda_{\min\{n,p'\}}$ and corresponding model parameters
$\{ (\hat{\theta}_o(\lambda_j), \hat{\boldsymbol{\theta}}(\lambda_j) ) \}_{j=1}^{\min\{n,p'\}}$
associated with different model dimensions $m\in\{1,\ldots, p'\}$.

\subsubsection{The LHS search}
\begin{figure}
    \centering
    \includegraphics[width=0.49\textwidth]{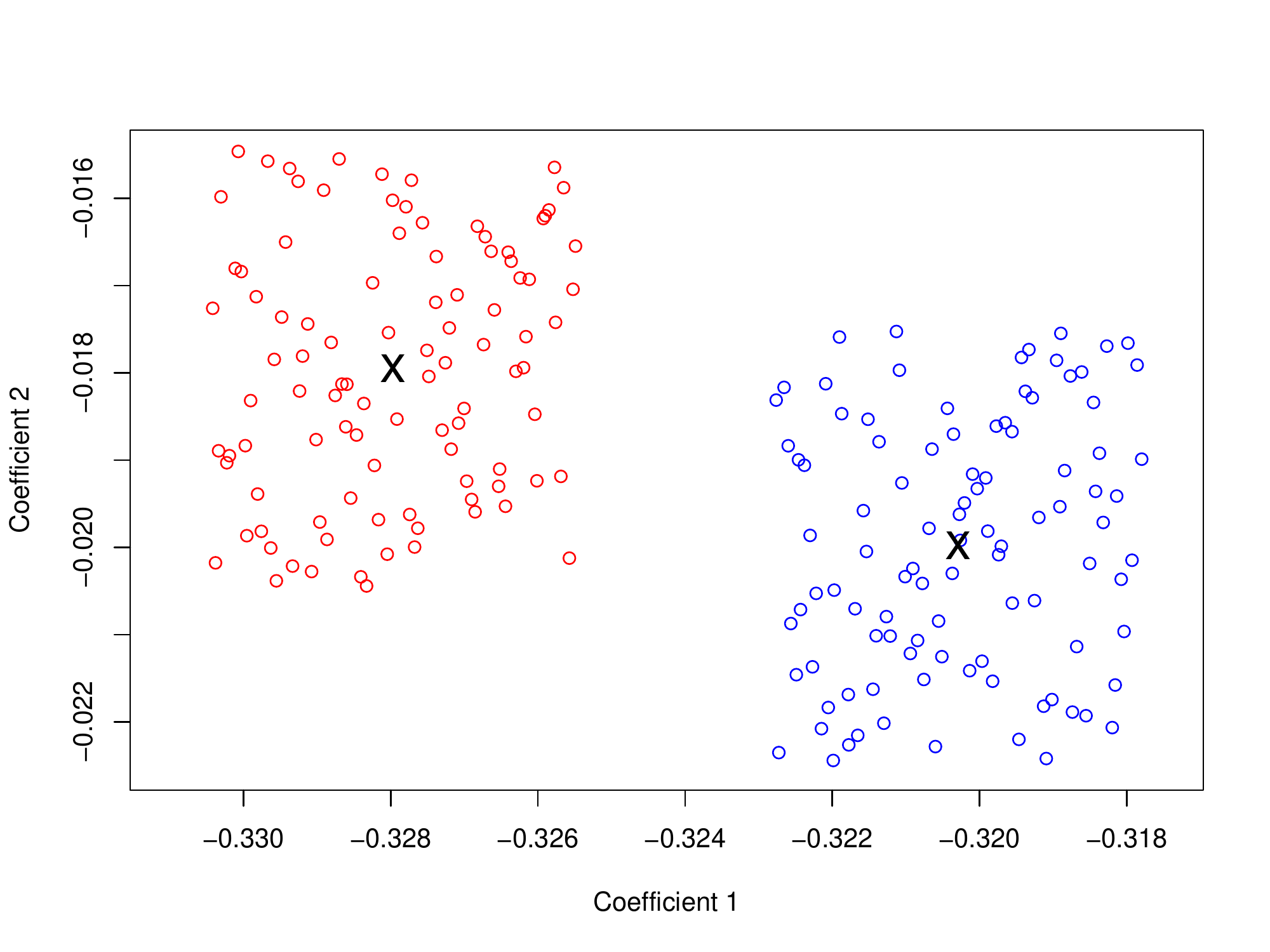}
    \includegraphics[width=0.49\textwidth]{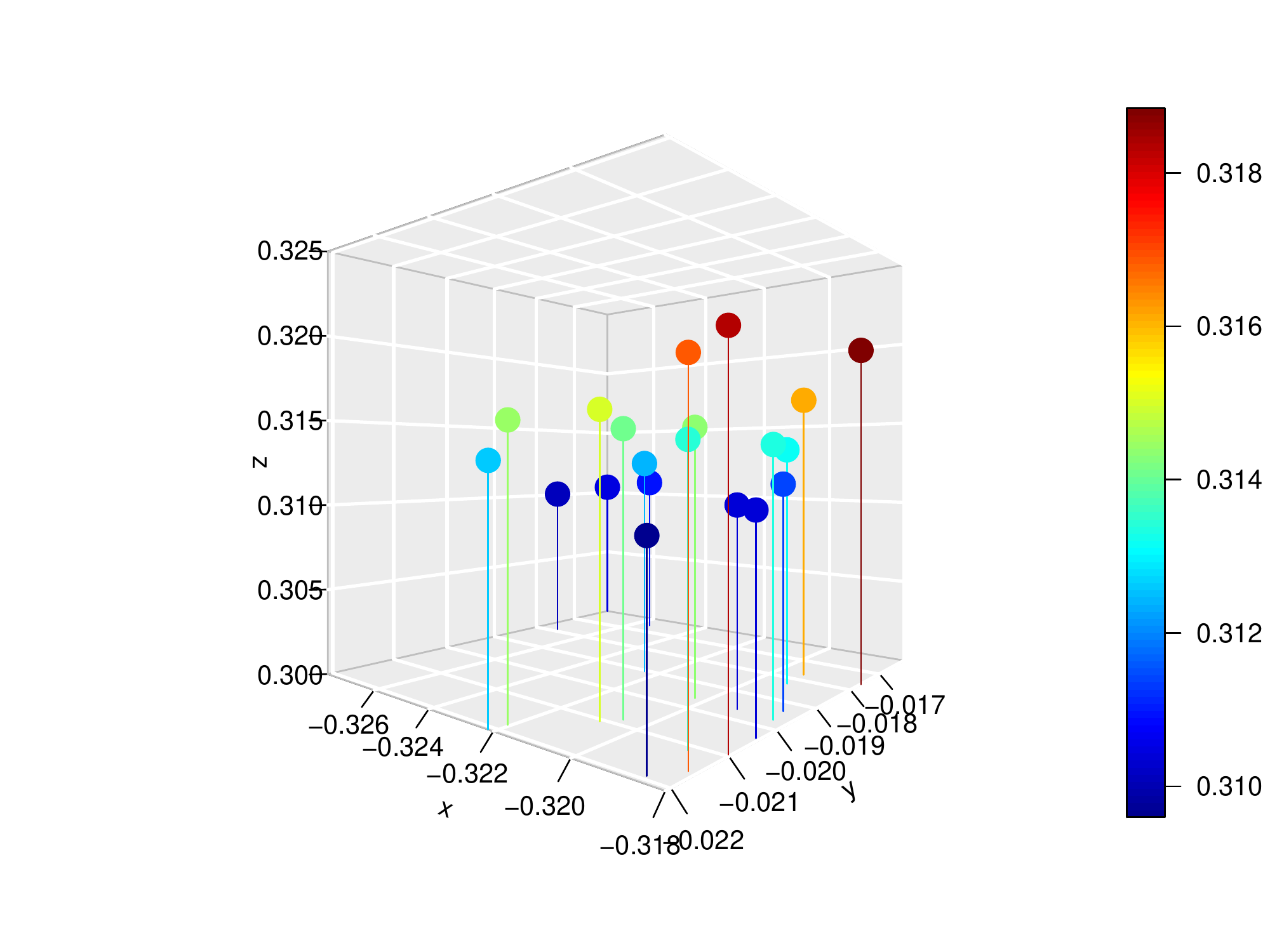}
    \caption{Example of Latin hypercube sampling around two different estimates $\hat{\boldsymbol{\theta}}(\hat{\lambda})$ for two coefficients. The idea is to sample points (left panel) and compute $\hat q_{\mathrm{adj}}$ at these points in order to maximize directly the adjusted Qini coefficient (right panel).}
    \label{fig:LHS_example}
\end{figure}

For each $\lambda_j$, $j=1,...,\min\{n,p'\}$,
we generate a LHS comprising $L$ points $\{\hat{\boldsymbol{\theta}}(\hat{\lambda}_j)_l\}_{l=1}^L$ in the neighborhood of $\hat{\boldsymbol{\theta}}(\hat{\lambda}_j)$,
and evaluate the adjusted Qini on each of these points.
The optimal coefficients 
are estimated as those coefficients among the $(\min\{n,p'\} \times L)$ LHS points
that maximize the adjusted Qini.
Figure~\ref{fig:LHS_example} illustrates the procedure.

\subsubsection{A simpler estimate of the Qini-based uplift regression parameters}
\label{sec:simpler:pur}

We also consider a simpler two-stage procedure to find a good uplift model.
This one is based only on the penalized log-likelihood and does not require the
posterior LHS-based search for the optimal coefficients.
Let $\hat q_{\mathrm{adj}}(\lambda )$ be the adjusted Qini coefficient associated with the model with parameters
$(\hat{\theta}_o(\lambda), \hat{\boldsymbol{\theta}}(\lambda))$.
The first stage of the procedure solves
\begin{align}
   \hat{\lambda} & =  \argmax \bigl( \hat q_{\mathrm{adj}}(\lambda_j) : j=1,\ldots, \min \{n,p'\} \bigr),
   \label{eq:firststage}
\end{align}
where as before, the sequence $\lambda_1 < \cdots < \lambda_{\min\{n,p'\}}$ is the logistic-lasso sequence.
On the second-stage, a reduced model that only include those explanatory variables associated with non-zero entries of the estimated parameter $\hat{\boldsymbol{\theta}}(\hat{\lambda})$ is fitted
without penalization, that is, with $\lambda$ set to zero.
The parameters are estimated with maximum likelihood.
This yields the selected model.
In our simulations, this model performs well.
It also serves to show that the value of
the penalization parameter $\hat \lambda$ that maximizes the Qini or
adjusted Qini, is not necessarily the same as the one that maximizes the penalized log-likelihood.

\section{Simulations}\label{sec:simulation}

We conduct a simulation study to examine the performance of Qini-based uplift regression. More specifically, we compare the different proposed parameters estimation methods by varying both the complexity of the data, and the number of predictors in the model.
In order to create realistic scenarios, we based our artificial data generation on the home insurance policy data described in the introduction. We take advantage of the opportunity to have real data in order to generate realistic scenarios. We proceed as follows. First, we fit a non-parametric model on a random sample $\mathcal{D}$ of the home insurance policy data. Based on the resulting model, we can extract the probabilities
$$p_1(\mathbf{x}) = \mathrm{Pr} (Y = 1 \mid \mathbf{x}, T = 1), \text{\ and\ } p_0(\mathbf{x}) = \mathrm{Pr} (Y = 1 \mid \mathbf{x}, T = 0),$$
for any given value  $\mathbf{x}$. Then, we use these probabilities to generate synthetic data. We start by creating a bootstrap sample $\mathcal{S}$ of size $n_\mathcal{S}$ from $\mathcal{D}$. For each observation $\mathbf{x}_i \in \mathcal{S}$, we generate a random vector $\tilde{y}_i =(\tilde{y}_{i0}, \tilde{y}_{i1})$, where $\tilde{y}_{i0}$ is the binary outcome of a Bernoulli trial with success probability $p_0(\mathbf{x}_i)$, and  $\tilde{y}_{i1}$ is the binary outcome of a Bernoulli trial with success probability  $p_1(\mathbf{x}_i)$, $i=1,\ldots, n_\mathcal{S}$. The augmented synthetic dataset $\{ (\x_i , t_i, \tilde{y}_i) \}_{i=1}^{n_{\mathcal{S}}}$, which we are going to denote again by $\mathcal{S}$, is the data of interest in the simulation. For each simulated dataset, we implement the following models:
\begin{enumerate}
    \item[(a)] a multivariate logistic regression without penalization as in (\ref{eq:uplift_likelihood}). This is the baseline model, and we will refer to it as \textit{Baseline}.
    \item[(b)] our Qini-based uplift regression model that uses several LHS structures to search for the optimal parameters (see Section~\ref{sec:qini:pur}). We denote this model by \textit{Q+LHS}.
    \item[(c)] our Qini-based uplift regression model that uses the simpler estimate
      of the regression parameters as explained in Section~\ref{sec:simpler:pur}.
      We denote this model by \textit{Q+lasso}. 
\end{enumerate}
Note that our \textit{Q+LHS} method is a derivative free optimization procedure.
Another derivative free optimization method is the well-known Nelder-Mead method \citep{nelder1965simplex}.
In order to obtain benchmarks for the LHS search, we implement the following Nelder-Mead Qini-based uplift regression models:
\begin{itemize}
\item[(d)] \textit{Base+NM}, which initializes the Nelder-Mead algorithm with the maximum likelihood estimates
  (the \textit{Baseline} model solution) and which searches for coefficients that maximize the adjusted Qini coefficient.
\item[(e)] \textit{Q+NM}, which initializes the Nelder-Mead algorithm with coefficients from the lasso-sequence
  (the first-stage of \textit{Q+lasso}) and which searches for coefficients that maximize the adjusted Qini coefficient.
\end{itemize}

\paragraph{Data generation}
As discussed in Section \ref{sec:literature2}, several tree-based methods have been suggested in the uplift literature. Here, we use the uplift random forest \citep{guelman2012random} as the data generating process. We chose this method due to its simplicity, and because it is readily
available in \textbf{R} through the package \textit{uplift} \citep{guelman2014uplift}. Algorithm \ref{alg:upliftRF} describes the associated methodology. 

\begin{algorithm}
\begin{algorithmic}[1]
  \Let{$B$}{number of bootstrap samples}
\For{$b=1$ to $B$}
\LongState{Draw a bootstrap sample of size $n_\mathcal{S}$ with replacement from the data}
\State{Fit an uplift decision tree $T_b$ to the bootstrap data}
\EndFor
\LongState{Output the ensemble of uplift trees $T_b$; $b=\{1,2,...,B\}$ and the predicted probabilities $\mathrm{Pr} (Y = 1 \mid \x, T = 1)$ and $\mathrm{Pr} (Y = 1 \mid \x, T = 0)$ obtained by averaging the predictions of the individual trees in the ensemble}

\end{algorithmic}
\caption{Uplift Random Forest \citep{guelman2012random}}
\label{alg:upliftRF}
\end{algorithm}

In our simulations, we vary two parameters: the depth of the trees used to fit the uplift random forests, and the number of variables $k$ considered when fitting the uplift logistic models. Algorithm \ref{alg:simulations} details the procedure.

We define $21$ scenarios by varying two parameters: (i) the depth of the uplift random forest trees used to generate the synthetic data is either $1$, $2$ or $3$, and (ii) the number of total covariates $k$ considered to build the forest model,
$k\in \{10, 20, 30, 50, 75, 90,97\}$. For scenarios 1-7, the depth is $1$,
and we vary $k$; for scenarios 8-14, the depth is $2$; and for scenarios 15-21, the depth is $3$.
Each scenario was replicated $100$ times.

\begin{algorithm}
\begin{algorithmic}[1]
\Let{$M$}{number of simulations}
\For{$m=1$ to $M$}
\LongState{Draw a stratified sample $\mathcal{D}_m$ of size $n_\mathcal{S}$ without replacement from the data}

\LongState{Fit an uplift random forest of a given tree depth to the sampled data $\mathcal{D}_m$} 

\LongState{Generate a complete (i.e., including the binary responses) synthetic data $\mathcal{S}_m$ with data $\mathcal{D}_m$}

\LongState{Fit an uplift random forest of the same given tree depth to the synthetic data $\mathcal{S}_m$ using all $p$ predictors}

\For{each $k$}
\State Sample $k \leq p$ random predictors for modeling
\State Fit the different uplift logistic models with $k$ predictors on $\mathcal{S}_m$
\EndFor
\EndFor
\LongState{Output the average and standard errors of the metrics for each method}
\end{algorithmic}
\caption{Simulations}
\label{alg:simulations}
\end{algorithm}

The sample means of $\hat q$ (\ref{eq:q_coeff}),
and $\hat q_{\mathrm{adj}}$ (\ref{eq:qadj}) and their corresponding standard errors are reported in Tables~\ref{tab:depth3_q},
and \ref{tab:depth3_qadj}, respectively.
Since the conclusions are similar for the three tree depths, we report only the results associated with depth $3$, that is, for the most complex model.
For each comparison group,  we also report the corresponding performance of an uplift random forest (\textit{RF}) fitted to the synthetic data using all available predictors (i.e., $k=97$).

\begin{table}[H]
    \centering
    \caption{Qini coefficient ($\hat q$) averaged over $100$ simulations. Standard-errors are reported in parenthesis. The \textit{RF} model (with $k=97$ and depth$=3$) performance is $1.60~(0.039)$. $n=5000$ observations.}
    \resizebox{\textwidth}{!}{
    \begin{tabular}{c|ccccc}
        \hline
         $k$ & Baseline & Q+lasso & Q+LHS & Base+NM & Q+NM  \\
         \hline
         $10$ & $0.51~( 0.023)$ & $0.56~( 0.020)$ & $0.76~( 0.021)$ &  $0.64~( 0.019)$ & $0.68~( 0.021)$\\
         $20$ & $0.74~( 0.020)$ & $0.83~( 0.022)$ & $1.03~( 0.024)$ & $0.93~( 0.025)$ & $0.95~( 0.020)$\\
         $30$ & $0.94~( 0.023)$ & $1.01~( 0.023)$ & $1.20~( 0.023)$ & $1.04~( 0.024)$ & $1.13~( 0.026)$\\
         $50$ & $1.09~( 0.039)$ & $1.19~( 0.025)$ & $1.48~( 0.026)$ & $1.23~( 0.031)$ & $1.35~( 0.029)$\\
         $75$ & $0.97~( 0.073)$ & $1.37~( 0.036)$ & $1.48~( 0.033)$ & $1.35~( 0.059)$ & $1.49~( 0.056)$\\
         $90$ & $1.00~( 0.084)$ & $1.36~( 0.045)$ & $1.54~( 0.036)$ &  $1.40~( 0.071)$ & $1.47~( 0.064)$\\
         $97$ & $0.83~( 0.083)$ & $1.36~( 0.055)$ & $1.59~( 0.037)$ &  $1.31~( 0.088)$ & $1.41~( 0.079)$\\
         \hline
    \end{tabular}}
    \label{tab:depth3_q}
\end{table}

In Table~\ref{tab:depth3_q}, we compare the performance of the models according to the Qini coefficient $\hat q$. We observe that performing variable selection driven by the adjusted Qini coefficient (\textit{Q+lasso}) significantly improves the performance of the baseline model. As expected, the models using a LHS-driven optimization perform better. The performance of \textit{Q+lasso} is similar to the \textit{Base+NM} performance, and is slightly lower than the others.
Using the lasso-sequence in order to initialize the posterior searches (\textit{Q+LHS} or \textit{Q+NM}) improves the performance of the final models. However, for the \textit{Q+NM} solution, the standard error of the Qini coefficient increases with $k$. 
It is almost twice the standard errors from \textit{Q+LHS} for $k \geq 75$.
Using all predictors ($k=97$) enables the \textit{Q+LHS} models to achieve the same performance as the \textit{RF} model.
\begin{table}[H]
    \centering
    \caption{Adjusted Qini coefficient ($\hat q_{\mathrm{adj}}$) averaged over $100$ simulations. Standard-errors are reported in parenthesis. The \textit{RF} model (with $k=97$ and depth$=3$) performance is $1.40~( 0.048)$. $n=5000$ observations.}
    \resizebox{\textwidth}{!}{
    \begin{tabular}{c|ccccc}
            \hline
          $k$ & Baseline & Q+lasso & Q+LHS & Base+NM & Q+NM  \\
          \hline
          $10$ & $0.36~( 0.027)$ & $0.39~( 0.026)$ & $ 0.72~( 0.024)$ & $0.54~( 0.025)$ & $0.61~( 0.027)$ \\
          $20$ & $0.58~( 0.025)$ & $0.68~( 0.027)$ & $ 1.02~( 0.025)$ & $0.83~( 0.025)$ & $0.91~( 0.027)$\\
          $30$ & $0.83~( 0.026)$ & $0.92~( 0.029)$ & $ 1.20~( 0.023)$ & $1.03~( 0.029)$ & $1.13~( 0.029)$\\
          $50$ & $0.97~( 0.041)$ & $1.12~( 0.026)$ & $ 1.48~( 0.026)$ & $1.23~( 0.033)$ & $1.35~( 0.028)$\\
          $75$ & $0.90~( 0.069)$ & $1.32~( 0.039)$ & $1.48~( 0.033)$ & $1.35~( 0.061)$ & $ 1.49~( 0.058)$\\
          $90$ & $0.94~( 0.081)$ & $1.28~( 0.049)$ & $ 1.54~( 0.036)$ & $1.37~( 0.071)$ & $1.46~( 0.069)$\\
          $97$ & $0.79~( 0.085)$ & $1.30~( 0.063)$ & $ 1.59~( 0.037)$ & $1.23~( 0.089)$ & $1.32~( 0.081)$\\
          \hline
    \end{tabular}
    }
    \label{tab:depth3_qadj}
\end{table}

In Table~\ref{tab:depth3_qadj}, we compare the main statistic of interest, that is, the adjusted Qini coefficient.
These results corroborate the findings from the previous table.
Guiding the variable selection by this statistic leads to significant improvements from the results of the baseline model.
Similarly, estimating the parameters with a derivative-free maximization of the adjusted Qini coefficient
improves the performance of the models in comparison to maximum likelihood estimation.
The best results are obtained with models that make use of the lasso-sequence in order to explore the space of the parameters
(\textit{Q+LHS} and \textit{Q+NM}). As in Table~\ref{tab:depth3_q},
the \textit{Q+LHS} models give the best results.
Moreover, when using all available predictors, the \textit{Q+LHS} models outperform both the \textit{RF} and the \textit{Q+NM} models.

\begin{figure}[H]
    \centering
    \includegraphics[width=0.48\textwidth]{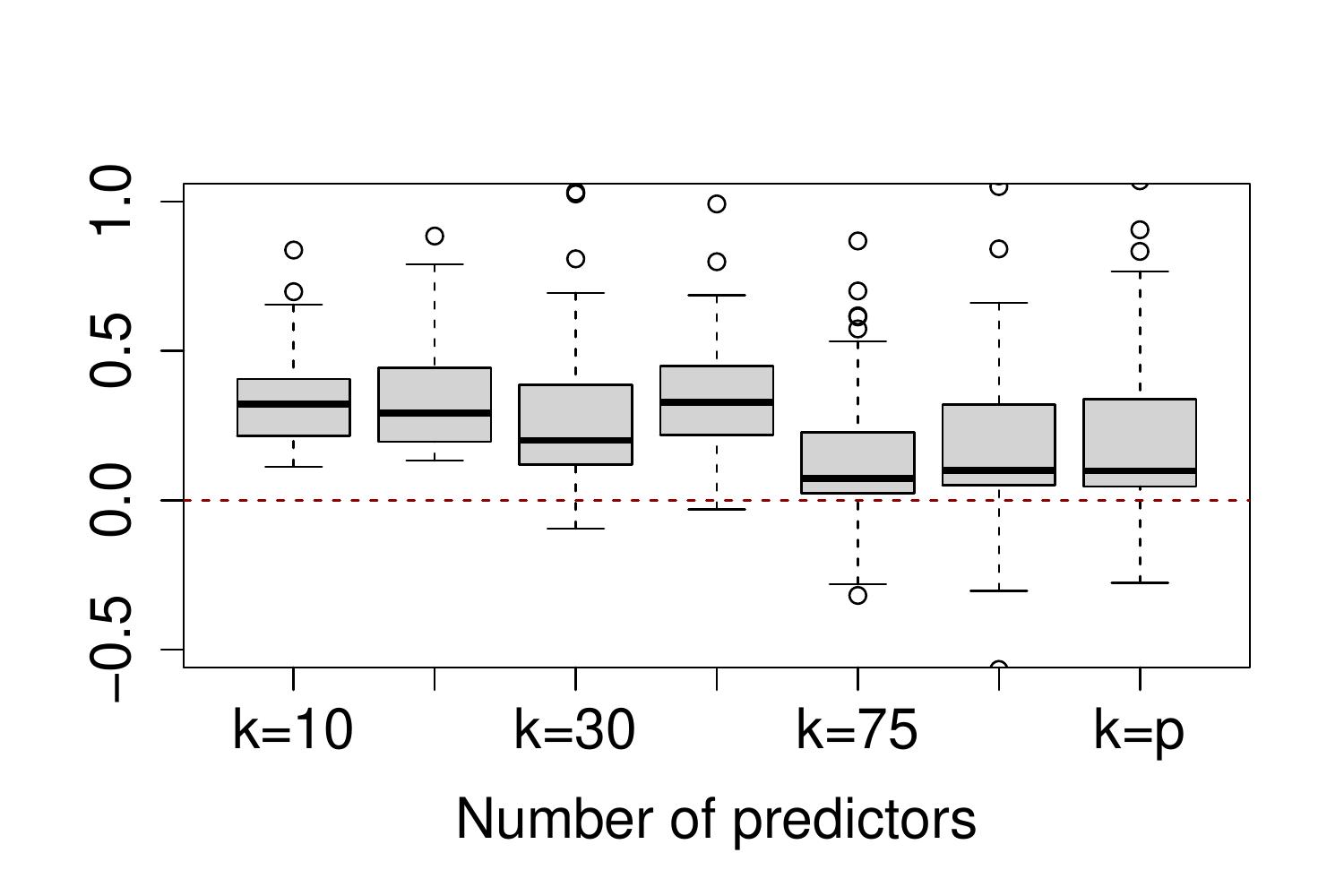}
    \includegraphics[width=0.48\textwidth]{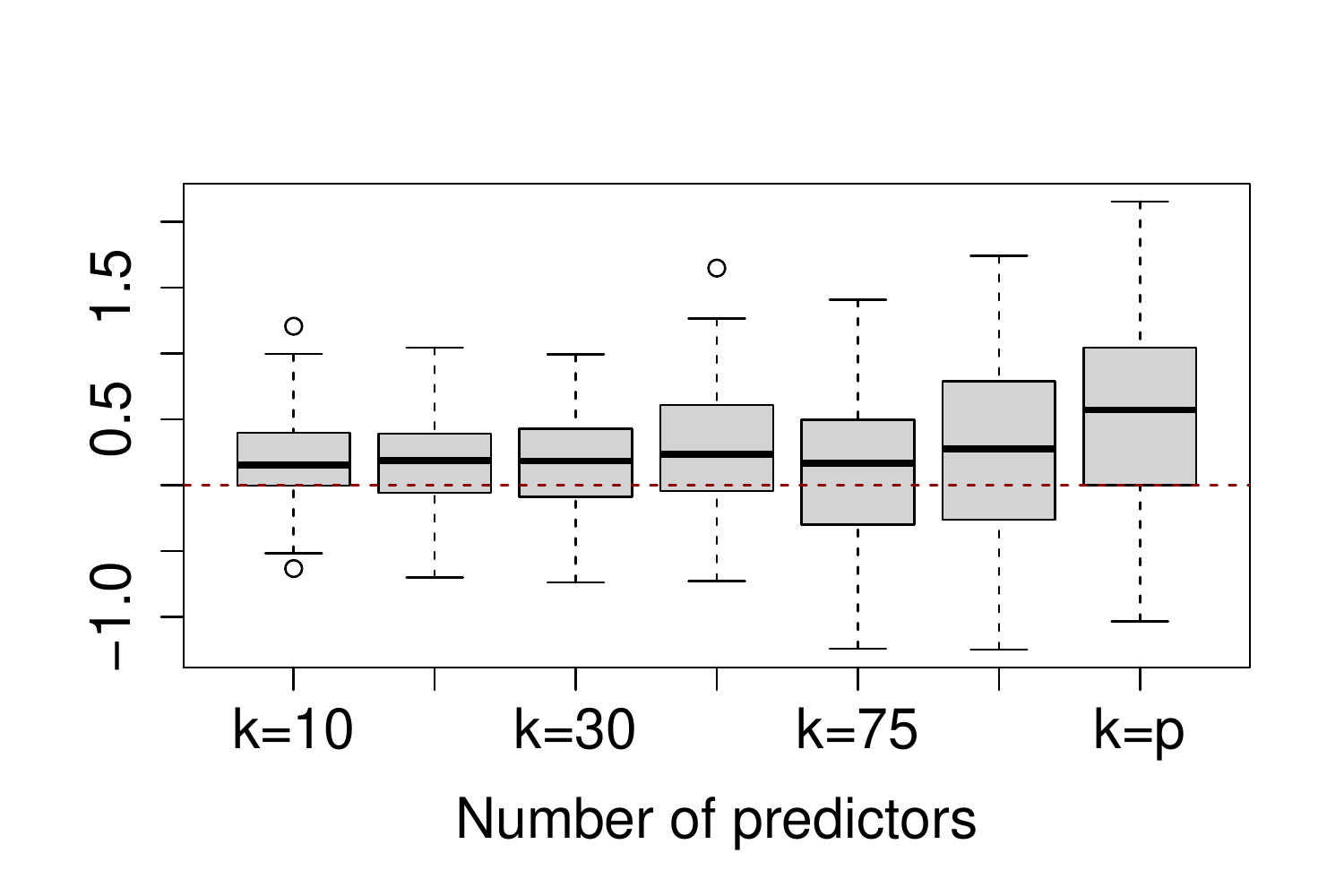}
    \caption{Comparison between \textit{Q+LHS} and \textit{Q+lasso} (left panel) and \textit{Q+LHS} and \textit{Base+NM} (right panel). Boxplots of the differences in terms of $\hat{q}_{\mathrm{adj}}$ as a function of the number of predictors used in the models over the $100$ simulations with $n=5000$ observations. The black lines represent the differences medians.}
    \label{fig:diff:lasso:LHS}
\end{figure}

The difference in performance between \textit{Q+LHS} and \textit{Q+lasso} is significant.
The left panel of Figure~\ref{fig:diff:lasso:LHS} display boxplots of the differences in performance between these two models
in each simulation. It is clear that \textit{Q+LHS} perform much better than \textit{Q+lasso} most of the time.
The relative performance of \textit{Q+lasso} improves slightly
when the number of predictors approaches the total number of predictors available ($p=97$).
The same pattern is observed in the difference of performance between \textit{Q+LHS} and \textit{Base+NM} (see right panel of Figure~\ref{fig:diff:lasso:LHS}).
This confirms the importance of the use of the lasso-sequence in order to estimate the model's parameters.

\begin{figure}[H]
    \centering
    \includegraphics[width=1\textwidth]{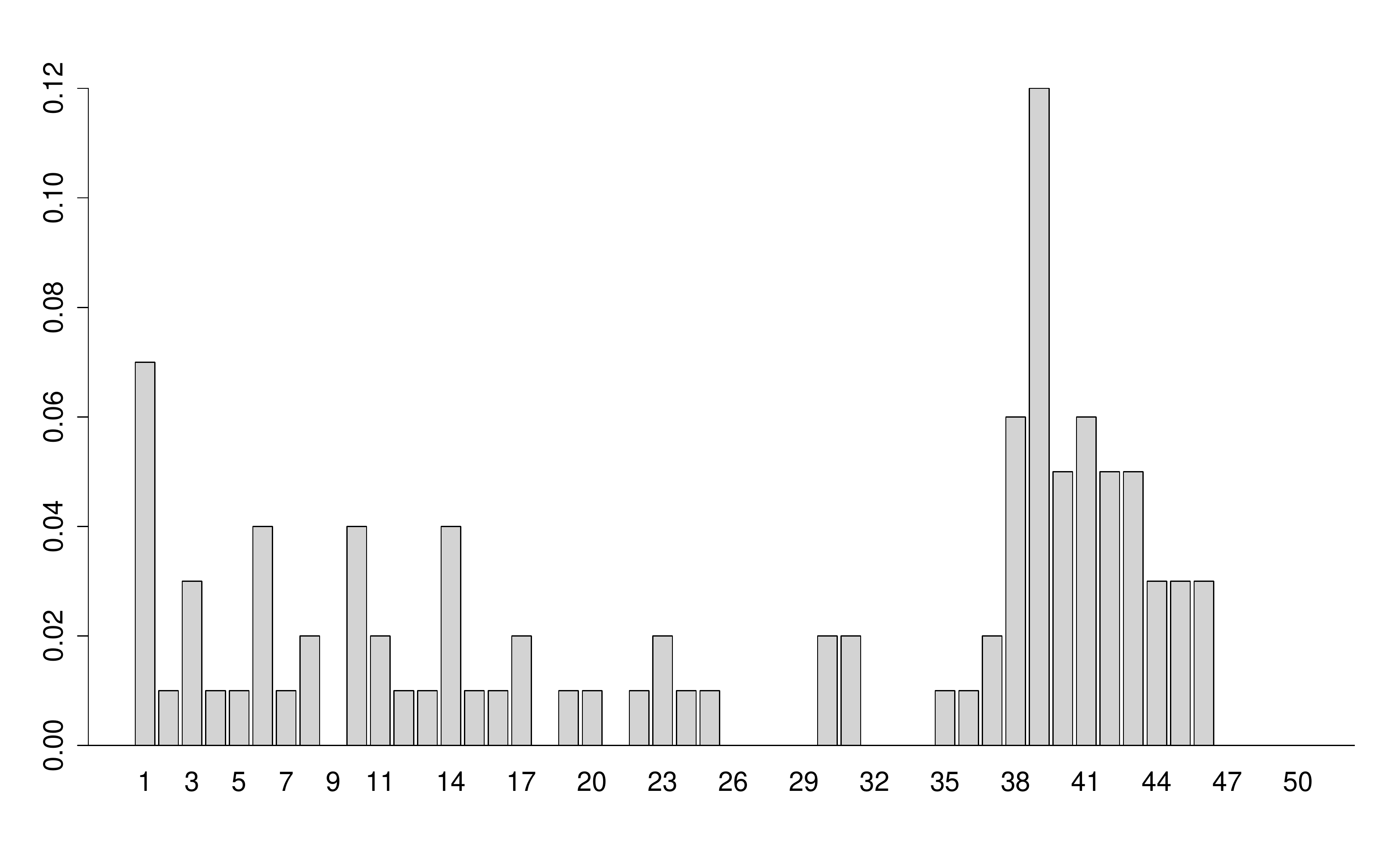}
    \caption{Barplot of the distribution of the Q+lasso rankings associated with $\tilde \lambda$.
    }
    \label{fig:loglik_ranking}
\end{figure}

\paragraph{Choosing an appropriate sparse model}
Next, we compare the models selected by
the Qini-based uplift regression and the classical \textit{lasso} approach where the penalization constant is chosen by cross-validation on the log-likelihood.
Consider the values of the logistic-lasso sequence $\lambda_1 <\cdots < \lambda_{\min\{n,p'\}}$ sorted according
to the results of \textit{Q+lasso}. That is, consider the permutation
$(\pi_1,\pi_2, \ldots, \pi_{\min\{n,p'\}})$ of $(1,2,\ldots, \min\{n,p'\})$ so that
$\lambda_{\pi_{\min\{n,p'\}}} \preceq \cdots \preceq \lambda_{\pi_1}$,
where the relation $\lambda_{\pi_i} \preceq \lambda_{\pi_j}$ means that
$\hat{q}_{\mathrm{adj}}( \lambda_{\pi_i}) \leq \hat{q}_{\mathrm{adj}}( \lambda_{\pi_j})$.
We look at the value of $\tilde \lambda \in\{\lambda_1,\ldots, \lambda_{\min\{n,p'\}}\}$
 that is chosen by cross-validation of the log-likelihood,
and report its ranking based on the sorted
Q+lasso sequence $\lambda_{\pi_{\min\{n,p'\}}} \preceq \cdots \preceq \lambda_{\pi_1}$.
Comparing the two models is equivalent to check when lasso
finds the ``best'' $\lambda$,
that is, when $\lambda_{\pi_1}$ is equal to $\tilde \lambda$.
We repeated the simulation $100$ times, each time using $n=5000$ observations randomly selected from the full data set.
The barplot in Figure~\ref{fig:loglik_ranking} shows that only $7\%$ of the time $\tilde \lambda$
also maximizes $\hat q_{\mathrm{adj}}$.
Observe that $4\%$ of the time $\tilde \lambda$ is positioned $10$ in the ranking, and
$12\%$ of the time, it is positioned $39$.
These results clearly show that choosing the penalization constant by
cross-validation of the log-likelihood
does not solve the problem of maximizing the adjusted Qini coefficient,
and therefore, is not necessarily appropriate for uplift models.

\section{Insurance data analysis}\label{sec:analysis}

Recall the insurance data introduced in Section \ref{sec:data}. The insurance company is interested in designing retention strategies to minimize its policyholders' attrition rate. An experimental loyalty campaign was implemented, from which policies coming up for renewal were randomly allocated into one of the following two groups: treatment group, and control group. The goal of this section is to analyze the marketing campaign results so as to identify both the set of persuadable clients, and the set of clients that should not be disturbed.
Table~\ref{tab:checkbalance} describes some of the $p=97$ available explanatory variables in the dataset, in addition to the treatment (Called or Control) and outcome (Renewed or Cancelled the policy) variables.

\begin{table}[H]
    \centering
    \caption{Descriptive statistics of some available variables for $n=20,997$ home insurance policies. With randomization, the difference of means between treatment and control groups is significantly not different from $0$ for all available predictors. For privacy concerns, we hide some values with *.}
    \begin{tabular}{lccccc}
        \hline
        & Control & Called & Diff Mean & Diff SD & Domain    \\ 
        \hline \\
        \bf{Sample size} & $18,672$ & $2,325$ &  &  &  \\
        \hline \\
        \bf{Credit Score} & 756.93 & 756.92 & -0.00 & 1.46 & $\mathbb{R}^+$\\ 
        \bf{Age (Years)} & 44.97 & 45.26 & 0.30 & 0.25 & $\mathbb{R}^+$\\ 
        \bf{Genger} &&&&&\\
        Male & 0.59 & 0.60 & 0.01 & 0.01 & $\{0, 1\}$\\ 
        \bf{Marital Status} &&&&&\\
        Divorced & 0.02 & 0.02 & 0.00 & 0.00 & $\{0, 1\}$\\ 
        Married & 0.69 & 0.69 & 0.00 & 0.01 & $\{0, 1\}$\\ 
        Single & 0.23 & 0.23 & 0.00 & 0.01 & $\{0, 1\}$\\ 
        \bf{Seniority (Years)} & 9.57 & 9.73 & 0.16 & 0.16 & $\mathbb{R}^+$\\
        \bf{Policy Premimm (\$)} &&&&&\\
        New Premium  & * & * & -7.44 & 16.14 & $\mathbb{R}^+$\\ 
        Old Premium  & * & * & -5.18 & 15.12 & $\mathbb{R}^+$\\ 
        \bf{Territory} &&&&&\\
        Rural & 0.06 & 0.06 & 0.00 & 0.01 & $\{0, 1\}$\\ 
        \bf{Products} &&&&&\\ 
        Auto and Home & 0.86 & 0.85 & -0.01 & 0.01 & $\{0, 1\}$\\ 
        \bf{Auto Policies Count} & 1.05 & 1.04 & -0.01 & 0.01 & $\mathbb{N}$\\ 
        \bf{Mortgage Count} & 0.66 & 0.67 & 0.01 & 0.01 & $\mathbb{N}$\\ 
        \bf{Residences Count} & 1.07 & 1.08 & 0.01 & 0.01 & $\mathbb{N}$\\
        \bf{Endorsement Count} & 1.98 & 2.00 & 0.02 & 0.03 & $\mathbb{N}$\\ 
        \bf{Neighbourhood's Retention} & 0.87 & 0.87 & 0.00 & 0.00 & $[0,1]$\\ 
        \bf{Type of Dwelling} &&&&&\\
        Family House & 0.69 & 0.70 & 0.00 & 0.01 & $\{0, 1\}$\\ 
        Duplex & 0.03 & 0.03 & -0.00 & 0.00 & $\{0, 1\}$\\ 
        Apartment & 0.19 & 0.19 & -0.00 & 0.01 & $\{0, 1\}$\\ 
        \bf{Year of Construction} & 1982.82 & 1983.52 & 0.71 & 0.58 & $\mathbb{N}$\\ \bf{Extra Options} &&&&&\\
        Option 1 & 0.20 & 0.20 & 0.00 & 0.01 & $\{0, 1\}$\\ 
        Option 2 & 0.73 & 0.73 & 0.00 & 0.01 & $\{0, 1\}$\\ 
        Option 3 & 0.71 & 0.72 & 0.01 & 0.01 & $\{0, 1\}$\\ 
        \hline
    \end{tabular}
    \label{tab:checkbalance}
\end{table}

\subsection*{Parameters estimation}

We fit the Qini-based uplift regression \textit{Q+LHS} to the data
using the methodology described in the previous sections.
For comparison purposes, we also considered the model \textit{Q+lasso}.
Although, we are interested in interpretable parametric models,
we also fit an uplift random forest (\textit{RF}) as a benchmark for our comparison. 
  
In order to choose the optimal value from the logistic-lasso sequence of penalization constant values $\{\lambda_1, \ldots, \lambda_{\min\{n,p'\}}\}$, we use
a $5$-fold cross-validation on the adjusted Qini statistics.
We compare the resulting models with the one yielded by applying the
classical lasso approach, that is, with the model associated with the value
of the penalization constant that maximizes the cross-validated log-likelihood. We will refer to this latter model as \textit{MLE+lasso}.
The two-stage approach was used in all the cases.
The first stage estimates the best $\lambda$ in the logistic-lasso sequence by cross-validation.
The second stage fits the non penalized logistic regression model with the subset
of selected variables.

For the \textit{Q+LHS} model, for each $\lambda_j$, we perform a LHS search to directly maximize the adjusted Qini coefficient. In this case, applying the LHS search leads to the selection of the model associated with the penalization constant $\approx 3 \times 10^{-5}$, while for the \textit{MLE+lasso} logistic regression, it is $\approx 10^{-3}$. The number of selected variables are, respectively, $163$ and $53$ out of a total of $195$ main and interaction effect terms. If we use the simple search method described in Section~\ref{sec:simpler:pur}, which was denoted by \textit{Q+lasso} in the previous section, the optimal value of the penalization constant is $\approx 4\times 10^{-5}$. In this case, the number of selected variables is $156$.

In order to have a fair comparison, we followed a process similar to the one applied to the \textit{Q+LHS} model to fit the uplift \textit{RF} model. The accuracy of a random forest can be sensitive to several training hyper-parameters: number of trees (from $10$ to $200$, with increments of  $10$ trees), maximum depth on each tree (from $1$ to $10$), minimum number of observations per node (either $100$, $200$ or $500$), and split criterion, either Euclidean distance or Kullback-Leibler divergence; see \cite{guelman2012random} for more details on the split criteria. The optimal \textit{RF} hyper-parameters were those that maximized the adjusted Qini coefficient with a $5$-fold cross-validation over the  grid given by the possible values of the hyper-parameters. Hence, the chosen \textit{RF} was composed of $100$ trees of maximum depth $3$, with a minimum of $200$ observations per node, with trees splitted according to the Kullback-Leibler criterion. The final \textit{RF} was  fitted using all available data.
  
\begin{figure}[!ht]
    \centering
    \includegraphics[width=1\textwidth]{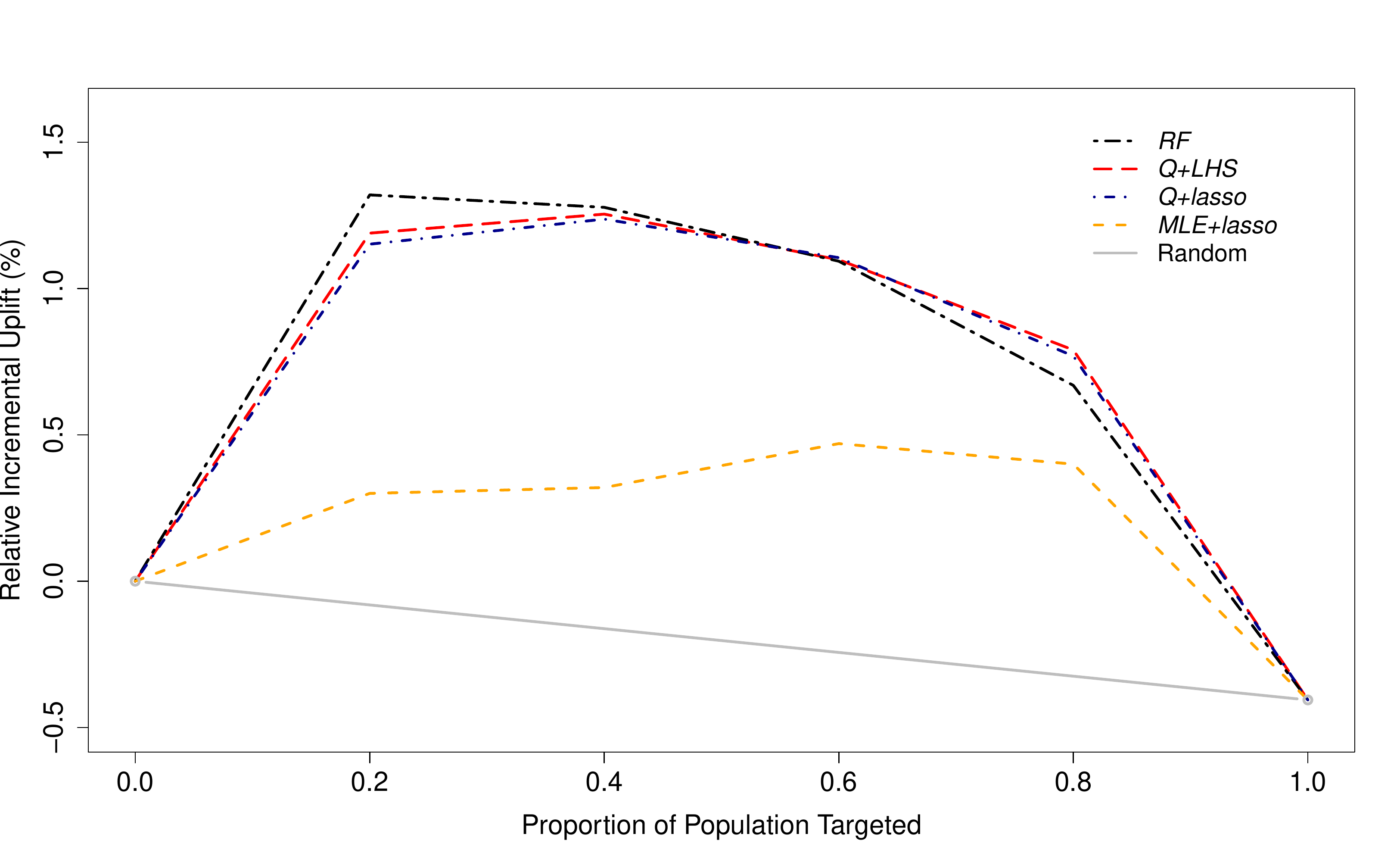}
    \caption{Performance of the final models based on the Qini curves.}
    \label{fig:qinicomp}
\end{figure}

\begin{figure}[!ht]
    \centering
    
    \includegraphics[width=0.49\textwidth]{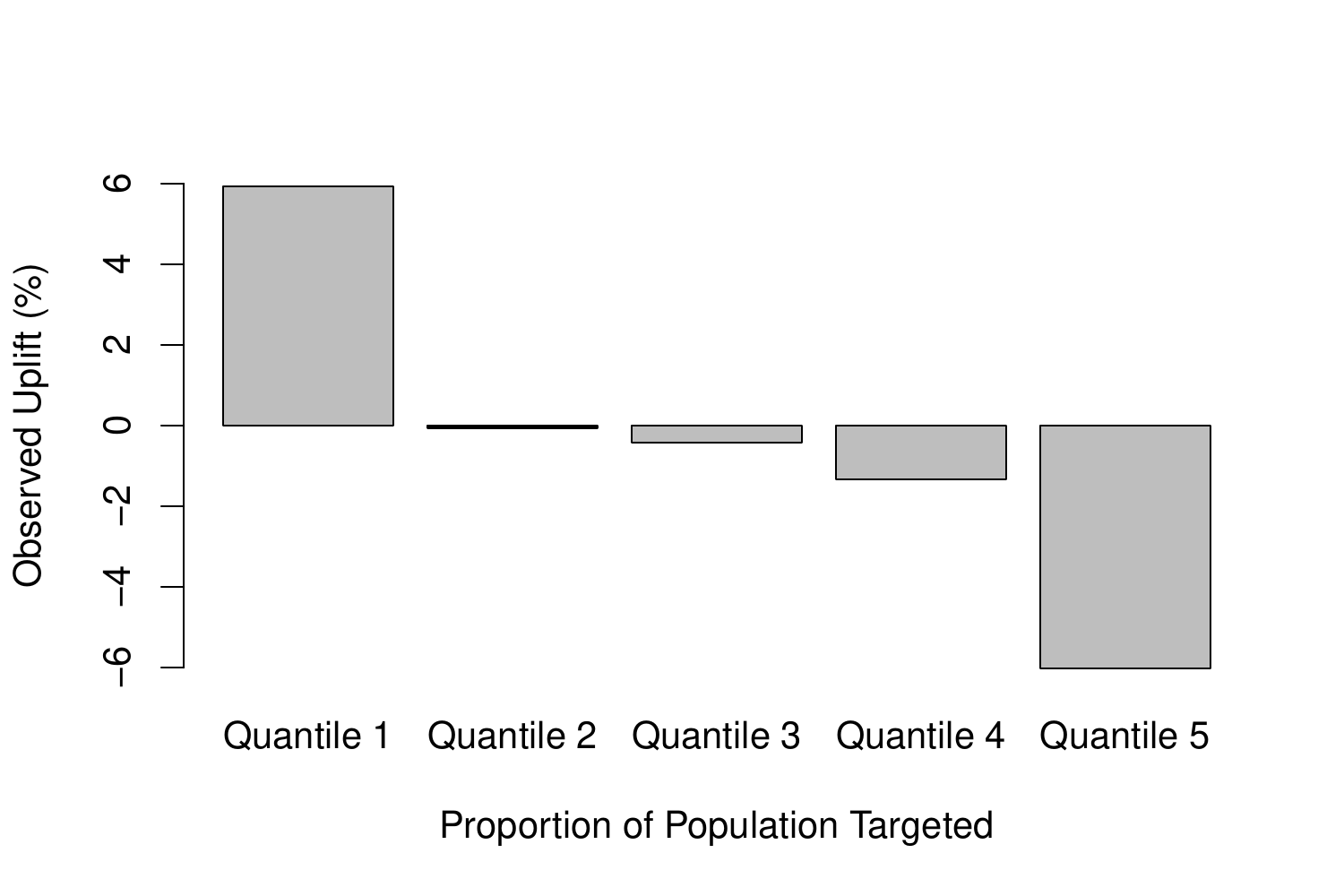}
    \includegraphics[width=0.49\textwidth]{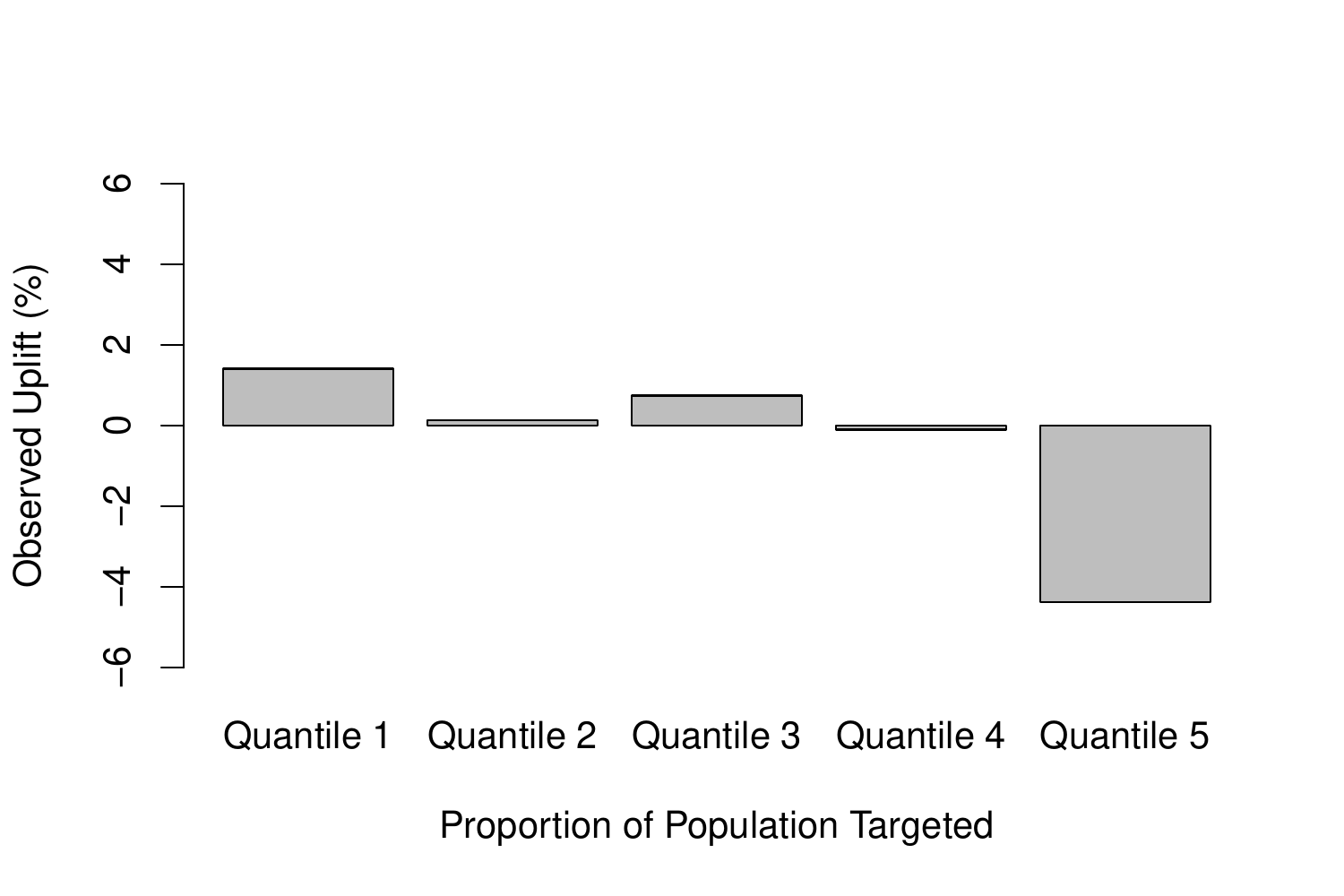}
     \caption{Performance of the Q+LHS and MLE+lasso models based on uplift Kendall's correlations. The left barplot corresponds to Q+LHS ($\rho=1$) and the right barplot to to the MLE+lasso model ($\rho=0.8$).}
    \label{fig:modelperf:kendall_qlhs_mlelasso}
\end{figure}

Figures~\ref{fig:qinicomp} and \ref{fig:modelperf:kendall_qlhs_mlelasso} show the performance of the models in terms of the Qini curve and the uplift barplot (Kendall's rank correlation), respectively.
As expected, the Qini-based uplift regression models outperform the classic lasso approach, that is,
the \textit{MLE+lasso} model, both in terms of overall adjusted Qini coefficient
(see Table \ref{tab:ModelComparison}) and in terms of sorting the individuals in decreasing order of uplift
(see Figure~\ref{fig:modelperf:kendall_qlhs_mlelasso}).
Moreover, the performance of the Qini-based uplift regression is slightly lower, but comparable to the one of the \textit{RF} model ($\hat{q}_{\mathrm{adj}} =1.07$), even though the \textit{RF} is more complex. Indeed, with $100$ trees of depth $3$, the \textit{RF} model can both model non-linearity and interactions between covariates, which makes interpretation of the final \textit{RF} hard. However, the in-sample performance is similar to our models. This is interesting because it shows that it is possible to get powerful models without loosing interpretation when estimating the parameters with the adjusted Qini function.
Based on the final models, we can identify both (i) the group of clients at the top  $20\%$ of predicted uplifts, that is, the clients to pursue in the marketing, and (ii) the group  of clients at the bottom  $20\%$ of predicted uplifts, that is, the clients not to disturb with any marketing. The group of clients at the top  $20\%$ of predicted uplifts provides very strong return on investment cases when applied to retention activities. For example, by
only targeting the persuadable customers in an outbound marketing
campaign, the contact costs and hence the return per unit spend can be
dramatically improved \citep{radcliffe2011real}. 
We observe from Table~\ref{tab:ModelComparison} that the \textit{RF} model finds the highest top $20\%$ uplift group which presents an uplift of $6.41\%$, while the \textit{Q+LHS} model finds the lowest bottom $20\%$ uplift group which shows an uplift of approximately $-6\%$.
Note that the overall uplift is approximately $-0.5\%$. The performance of the \textit{Q+lasso} model is slightly lower than the one of the \textit{Q+LHS} model.
\begin{table}[!ht]
    \centering
    \caption{Uplift comparison of the top and bottom $20\%$ uplift groups estimated by the models: \textit{\textit{RF}, Q+LHS, Q+lasso,} and \textit{MLE+lasso.}}
    \begin{tabular}{l|ccc}
         \hline
         \bf{Method} & \bf{$\hat q_{\mathrm{adj}}$} & \bf{Top $20\%$ Uplift} & \bf{Bottom $20\%$ Uplift} \\
         \hline
         \textit{RF} & $1.07$ & $6.41\%$ & $-5.36\%$ \\
         \textit{Q+LHS} & $1.03$ & $5.94\%$ & $-6.02\%$ \\
         \textit{Q+lasso} & $1.02$ & $5.76\%$ & $-5.93\%$ \\
         \textit{MLE+lasso} & $0.39$ & $1.41\%$ & $-4.38\%$ \\
         \hline
    \end{tabular}
    \label{tab:ModelComparison}
\end{table}

\subsection*{Model interpretation}
Because the \textit{Q+LHS} model is a logistic model, we can interpret the results through its coefficients. The usual approach is that of the odds ratios. For a specific variable, the odds ratio is computed by fixing the other covariates at fixed values, such as their mean, which is what we have done here.
Since the company is not interested in all the variables included in the model, we will analyze a subset with relevant interpretation for the business. In addition, for confidentiality reasons, we do not show the analysis of variables related to the insurance premium. The following variables are chosen by our model: client's \textit{ credit score, age, gender} and \textit{ marital status} (single or not); client's \textit{products:} whether it is a single line (home) or a multi-line (automobile and home) account; client's number of \textit{automobile policies, mortgages} and \textit{residences}; and whether the client's has \textit{extra options} (additional endorsements) in his/her account. For a model with $p$ variables, the odds ratio $\mathrm{OR}_{X_j}(t)$ for a specific variable $X_j$ is given by 
\begin{align*}
    &\frac{\mathrm{Pr} (Y = 1 \mid X_j=x_j+1, T = t) / \mathrm{Pr} (Y = 0 \mid X_j=x_j+1, T = t)}{\mathrm{Pr} (Y = 1 \mid X_j=x_j, T = t) / \mathrm{Pr} (Y = 0 \mid X_j=x_j, T = t)}\\
    &=\frac{\mathrm{exp}(\hat\beta_j (x_j+1) + \hat\delta_j t~(x_j+1))}{\mathrm{exp}(\hat\beta_j x_j + \hat\delta_j t~x_j)} 
    =\mathrm{exp}(\hat\beta_j) ~ \mathrm{exp}(\hat\delta_j t) ~ ,
\end{align*}
where $T$ is the treatment indicator, and where for a binary variable, such as \textit{extra options}, $x_j$ is set to $0$ in the above expression.
When the company does not call a client ($T=0$), the odds ratio is $\mathrm{OR}_{X_j}(0) = \mathrm{exp}(\hat\beta_j)$ and when the company calls a client ($T=1$), the odds ratio is $\mathrm{OR}_{X_j}(1) = \mathrm{exp}(\hat\beta_j)~ \mathrm{exp}(\hat\delta_j) =\mathrm{OR}_{X_j}(0)~ \mathrm{exp}(\hat\delta_j)$. 
Table~\ref{tab:PUR1_odds} gives the estimated odds ratios $\mathrm{OR}_{X_j}(0)$ and $\mathrm{OR}_{X_j}(1)$ with $95\%$ confidence intervals.  We can see, for example,  that when the company does not call a client which has \textit{extra options} in his/her policy, his/her odds ratio of renewing the policy is $0.35$ while when the company calls that same client, the odds ratio becomes $1.28$.

\begin{table}[!ht]
    \centering
    \caption{Odds ratios and $95\%$ confidence intervals estimated by the Qini-based uplift regression model (\textit{Q+LHS}) for some of the selected variables, $^*$ represents significant coefficients.}
    \begin{tabular}{lcccc}
         \hline
         & $\mathrm{exp}(\hat\beta_j)$  & \bf{CI ($95\%$)} &  $\mathrm{exp}(\hat\beta_j+\hat\delta_j)$ &  \bf{CI ($95\%$)}  \\ 
    \hline \\
    \bf{Credit Score} & $0.998$  & $(0.994;~1.003)$ &  $1.001$ & $(0.999;~1.001)$\\
    \bf{Age (Years)} & $0.995$ & $(0.969;~1.023)$ &  $0.998$ & $(0.989;~1.006)$\\
    \bf{Genger} &&&&\\
    Male & $1.297$ & $(0.778;~2.163)$ &  $0.962$ & $(0.814;~1.135)$\\
    \bf{Marital Status} &&&&\\
    Single & $2.759$ & $(0.956;~7.963)$ &   $0.697$ & $(0.452;~1.077)$\\
    \bf{Products} &&&&\\ 
    Auto and Home & $1.619$ & $(0.586;~4.472)$ & $^*1.418$ & $(1.017;~1.977)$\\
    \bf{Auto Policies Count} & $2.106$ & $(0.653;~6.790)$ &  $^*1.996$ & $(1.368;~2.918)$\\
     \bf{Mortgage Count} & $1.381$ & $(0.789;~2.418)$  & $^*1.366$ & $(1.137;~1.642)$\\
     \bf{Residences Count} & $0.505$ & $(0.172;~1.489)$ &  $^*1.788$ & $(1.122;~2.848)$\\
     \bf{Extra Options} & $^*0.350$ & $(0.141;~0.874)$ &  $1.276$ & $(0.949;~1.715)$\\
     \hline
    \end{tabular}
    
    \label{tab:PUR1_odds}
\end{table}

\begin{figure}
    \centering
    \includegraphics[width=0.48\textwidth]{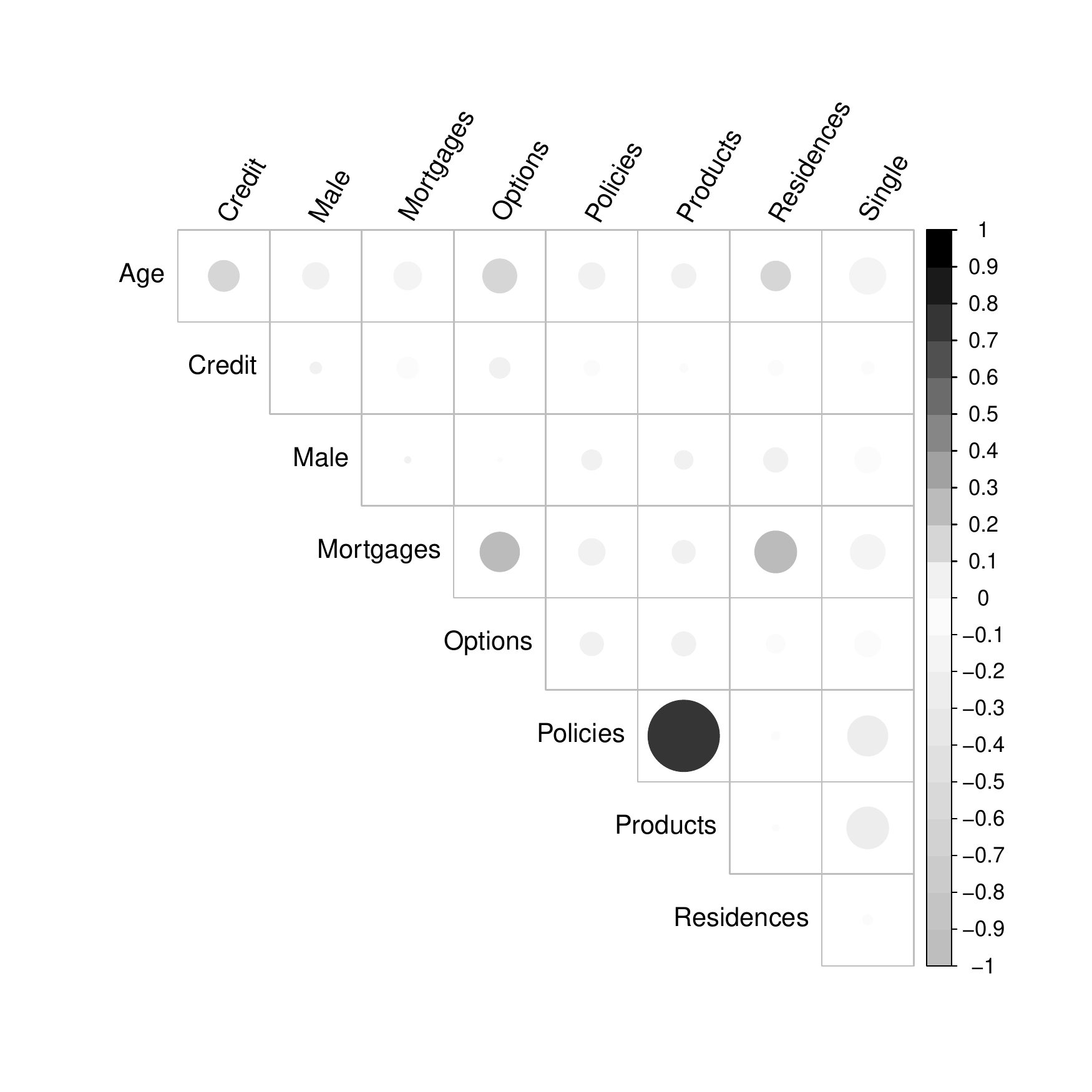}
    \includegraphics[width=0.48\textwidth]{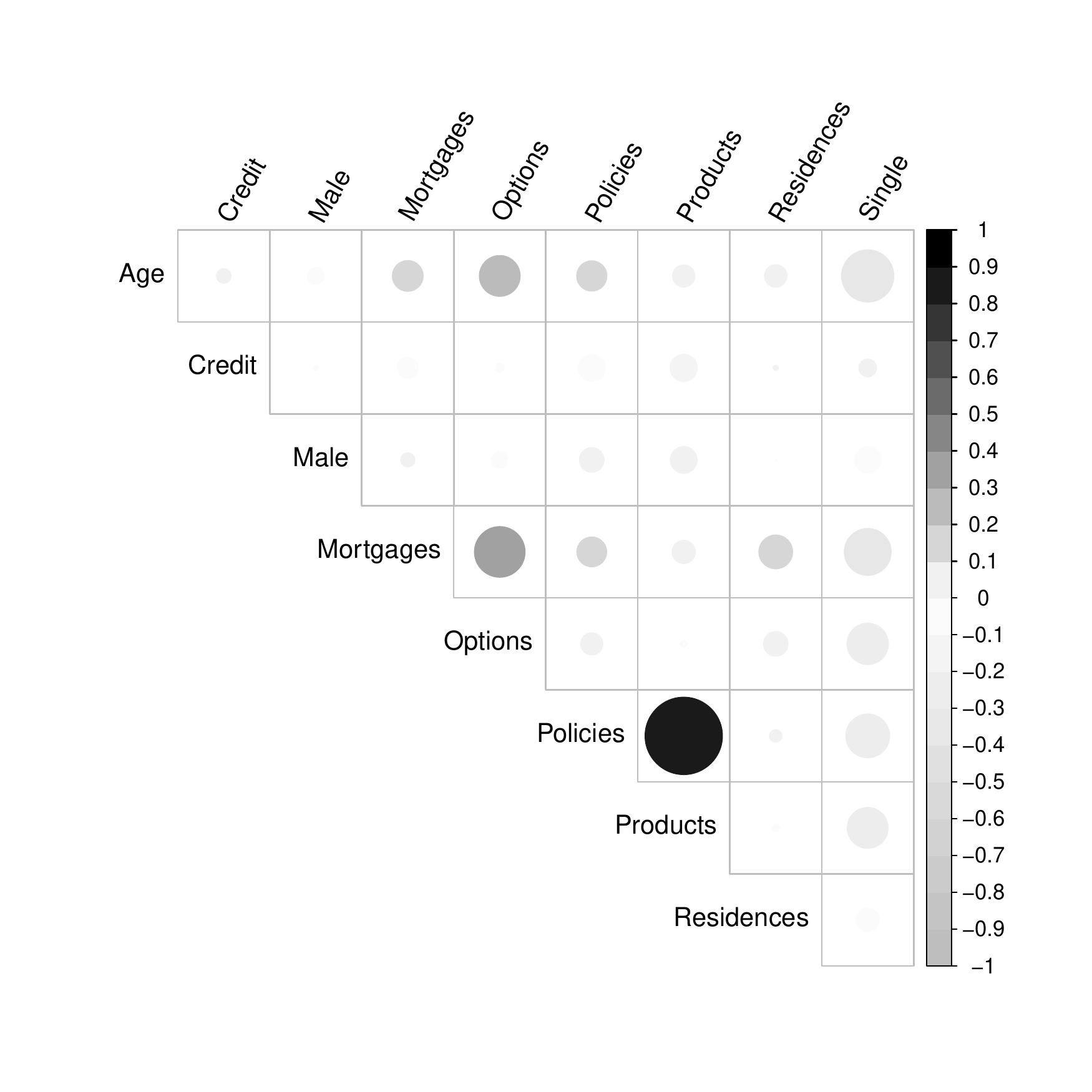}
    \caption{Correlations of selected variables of interest for Persuadable (left panel) and Do-not-disturb (right panel) clients.}
    \label{fig:correlations}
\end{figure}

Next, we use the \textit{Q+LHS} model predictions to describe in more detail the two extreme groups found by the model (top $20\%$ and bottom $20\%$ predicted uplifts). This furnishes the insurance company with typical profiles of clients that are persuadables (top $20\%$), and clients that should not be targeted (bottom $20\%$).
Table~\ref{tab:PUR1_profile} shows descriptive statistics
of some selected predictors for both groups.  A \textsc{manova} comprising only these two groups
for the selected variables, followed by \textsc{anova} tables involving individual selected variables separately, show that all mean differences were statistically significant ($p$-value $< 0.0005$).
\begin{table}[!ht]
\centering
\caption{Profiles of the persuadables and do not disturb groups predicted by the Qini-based uplift regression model (\textit{Q+LHS}) for some of the selected variables. Note that all group means are significantly different from 0 ($p$-value $< 0.0005$).}
\begin{tabular}{lcc}
    \hline
    & \bf{Persuadables}  & \bf{Do Not Disturb}   \\ 
    \hline \\
    \bf{Number of observations} & $4199$ & $4200$ \\
    \bf{Observed Uplift}  & $5.94\%$ & $-6.02\%$ \\
    \bf{Predicted Uplift} ($\pm$ S.E.)  & $4.94\%~(\pm 0.10\%)$ & $-5.45\%~(\pm 0.07\%)$ \\
    \hline \\
    \bf{Credit Score} & $771~(\pm 60)$ & $736~(\pm 77)$\\
    \bf{Age (Years)} & $46.1~(\pm 11.6)$ & $41.4~(\pm 11.8)$\\ 
    \bf{Genger} &&\\
    Male &  $50\%~(\pm 50\%)$ & $61\%~(\pm 49\%)$\\
    \bf{Marital Status} &&\\
    Single &  $13\%~(\pm 33\%)$ & $43\%~(\pm 49\%)$\\ 
    \bf{Products} &&\\ 
    Auto and Home & $83\%~(\pm 37\%)$ & $66\%~(\pm 47\%)$\\
    \bf{Auto Policies Count} & $1.04~(\pm 0.64)$ & $0.76~(\pm 0.63)$\\ 
    \bf{Mortgage Count} & $0.63~(\pm 0.63)$ & $0.51~(\pm 0.55)$\\ 
    \bf{Residences Count} & $1.15~(\pm 0.41)$ & $1.02~(\pm 0.19)$\\
    \bf{Extra Options} & $87\%~(\pm 33\%)$ & $41\%~(\pm 49\%)$ \\
    \hline
    \end{tabular}
    \label{tab:PUR1_profile}
\end{table}
Looking at the average profiles of \textit{persuadable} and \textit{do not disturb} clients, we can say that a \textit{persuadable} client has a higher credit score and is slightly older than a client that should not be targeted.
A \textit{persuadable} client  is less likely to be single and more likely to hold both company insurance products (i.e., home and auto policies). Also, this type of client holds more auto policies in his/her account, more mortgages, more residences in his/her name and is more likely to have extra coverage options. Thus, it seems that a persuadable client is a customer with many products to insure.
The correlation matrices associated with these two groups are displayed in image format in Figure~\ref{fig:correlations}. There are some obvious patterns that distinguish the two groups. For example, \textit{credit score} is slightly correlated with \textit{client age} for persuadable clients, but not for do-not-disturb clients. Client \textit{age} is negatively correlated with \textit{marital status} for do-not-disturb clients, but only slighlty correlated for persuadables. Indeed, there are several differences in the \textit{marital status} correlations in both groups. Also, the number of \textit{mortgages} and \textit{residences} are more correlated for persuadables than do-not-disturb, and  the number of \textit{mortgages} and whether or not a client has \textit{extra options} are more correlated for do-not-disturb than persuadables.

The differences between these two groups can also be observed through the odds ratios.
For any  specific variable $X_j$ which takes average values $x_j^{(\mathrm{p})}$ (for persuadable clients), and $x_j^{(\mathrm{d})}$ (for do-not-disturb clients), consider the odds ratio $\mathrm{OR}_{X_j}^{\mathrm{(group)}}(t)$ between persuadable and do-not-disturb clients
\begin{align}
    &\frac{\mathrm{Pr} (Y = 1 \mid X_j=x_j^{(\mathrm{p})}, T = t) / \mathrm{Pr} (Y = 0 \mid X_j=x_j^{(\mathrm{p})}, T = t)}{\mathrm{Pr} (Y = 1 \mid X_j=x_j^{(\mathrm{d})}, T = t) / \mathrm{Pr} (Y = 0 \mid X_j=x_j^{(\mathrm{d})}, T = t)} \nonumber\\
    &=\bigl(\mathrm{exp}(\hat\beta_j) ~ \mathrm{exp}(\hat\delta_j t)\bigr)^{x_j^{(\mathrm{p})} - x_j^{(\mathrm{d})}} = \bigl(\mathrm{OR}_{X_j}(t)\bigr)^{x_j^{(\mathrm{p})} - x_j^{(\mathrm{d})}}~,\label{eq:or}
\end{align}
where $T$ is the treatment indicator. Table~\ref{tab:PUR1_odds_per_vs_donotdist} shows these odds ratios for the two values of $T \in \{0,1\}$.
For example, if we only consider \textit{extra options,} when the insurance company calls a client (i.e., $T=1$), the odds ratio between a persuadable client (Extra Options$=87\%$) and a do-not-disturb client (Extra Options$=41\%$) is about $1.12$ with a $95\%$ confidence interval of $[0.98;~1.28]$. On the other hand, when the company does not call a customer (i.e., $T=0$), the odds ratio becomes $0.62$ with a $95\%$ confidence interval of $[0.41;~0.94]$.
These results are quite logical in the sense that the odds of renewing the insurance policy are higher for the persuadable clients if the company calls, while the same odds are higher for the do-not-disturb clients if the company does not call.
\begin{table}[!ht]
\centering
\caption{Odds ratios $\mathrm{OR}_{X_j}^{\mathrm{(group)}}(t)$ of the persuadable compared to the do not disturb clients (Eq.~\ref{eq:or}) and $95\%$ confidence intervals estimated by the Qini-based uplift regression model (\textit{Q+LHS}) for some of the selected variables. The $\Delta = {x_j^{(\mathrm{p})} - x_j^{(\mathrm{d})}}$ column represents the difference of group means from Table~\ref{tab:PUR1_profile}.}
\begin{tabular}{lccccc}
\hline
 & $\Delta$   & \bf{Control}  & \bf{CI ($95\%$)}  & \bf{Called} & \bf{CI ($95\%$)} \\ 
\hline \\
\bf{Overall} & -  & $0.52$ & $(0.26;~1.03)$ &  $1.81$ & $(1.43;~2.29)$\\
\hline \\
\bf{Credit Score} & $35$  & $0.95$ & $(0.82;~1.09)$ & $1.02$ & $(0.98;~1.07)$\\
\bf{Age (Years)} & $4.7$  & $0.98$ &  $(0.86;~1.11)$& $0.99$ & $(0.95;~1.03)$\\ 
\bf{Genger} &&&&&\\
Male & $-11\%$ & $0.97$ & $(0.92;~1.03)$ & $1.00$ & $(0.99;~1.02)$ \\ 
\bf{Marital Status} &&&&&\\
Single & $-30\%$  & $0.73$ & $(0.54;~1.01)$ & $1.11$ & $(0.98;~1.27)$\\ 
\bf{Products} &&&&&\\ 
Auto and Home & $17\%$  & $1.09$ & $(0.91;~1.29)$ & $1.06$ & $(1.00;~1.12)$\\ 
\bf{Auto Policies Count} & $0.28$ & $1.23$ & $(0.89;~1.71)$ & $1.21$ & $(1.09;~1.35)$ \\ 
\bf{Mortgage Count} & $0.12$ & $1.04$ & $(0.97;~1.11)$ & $1.04$ & $(1.02;~1.06)$\\ 
\bf{Residences Count} & $0.13$ & $0.92$ & $(0.80;~1.05)$ & $1.08$ & $(1.02;~1.15)$\\ 
\bf{Extra Options} & $46\%$ & $0.62$ & $(0.41;~0.94)$ & $1.12$ & $(0.98;~1.28)$\\ 
\hline
\end{tabular}
\label{tab:PUR1_odds_per_vs_donotdist}
\end{table}

Overall, we observe that when calling a client, the odds of renewing the insurance policy of persuadable clients are almost twice ($1.81$) the odds of do-not-disturb clients with a $95\%$ confidence interval of $[1.43;~2.29]$. Conversely, when the company does not call a client, the odds of renewing the insurance policy of persuadable clients are half ($0.52)$ the odds of do-not-disturb clients with a $95\%$ confidence interval of $[0.26;~1.03]$. Hence, based on our model, by calling identified persuadable clients and not calling identified do-not-disturb clients in future marketing campaigns should result in increased retention rates for the company.

\subsection*{Uplift prediction}
The main objective in analyzing the insurance data is to estimate the parameters of the parametric uplift model which maximizes the Qini. Based on these estimates, we were able to provide useful insights to the company. In order to prevent overfitting, we made use of $5$-fold cross validation in the fitting process. However, since uplift models can also be used for predicting future clients behaviour, it is important to evaluate out-of-sample performance. In Table~\ref{tab:ModelComparison}, we showed the in-sample performance. Since we do not have a test sample, in order to evaluate the out-of-sample performance, we proceeded in the following way. 
  We ran $30$ experiments. For each experiment, we reserved $25\%$ randomly drawn observations for out-of-sample performance (test-set). We used the remaining observations to fit the models. These observations were further randomly divided into
  training-set, which comprised $\nicefrac{2}{3}$ of the remaining observations, and validation-set.
  We compare two ways of fitting the models. First, we only use the training set to fit the models,
  and compute the test-set performance through the adjusted Qini coefficient; the validation-set was not use in this process.
  Second, as before, we use the training data to fit the models, but the model parameters and/or coefficients
  are chosen so as to find the best fit for the validation-set (cross-validation).
  For each experiment, the training-set size was $10,394$ observations, the validation-set size was $5,354$ observations, and the test-set size was $5,249$ observations.
  To fit the \textit{RF}s models, we searched for the hyper-parameters that maximize the adjusted Qini coefficient
  following the same procedure that was applied in the first part of the insurance data analysis.

  The results of the only training-set way of fitting the models are displayed in Table~\ref{tab:perf_train_only}.
  We see that the \textit{RF} model shows the highest performance in the training-set.
  However, there are strong signs of overfitting.
  The \textit{Q+LHS} models clearly outperform the other methods.

\begin{table}[H]
    \centering
    {
     \caption{Out-of-sample performance when models are trained using the training observations only. The adjusted Qini coefficients are averaged over $30$ experiments. Standard-errors are shown in parenthesis.}
    \begin{tabular}{l|cc}
             \hline
             \bf{Method} & \bf{training-set} & \bf{test-set} \\
             \hline
             \textit{RF} & $\bf 1.195~(0.020)$ & $0.048~(0.014)$ \\
             \textit{Q+LHS} & $0.993~(0.029)$ & $\bf 0.703~(0.060)$ \\
             \textit{Q+lasso} & $0.896~(0.025)$ & $0.093~(0.021)$ \\
             \textit{MLE+lasso} & $0.481~(0.033)$ & $0.033~(0.012)$ \\
             \hline
    \end{tabular}
    \label{tab:perf_train_only}}
\end{table}

In order to mitigate the overfitting seen in the experiments where the models were fit using only the training observations, we now choose the model that maximizes the adjusted Qini on the validation set. Then, we score the observations from the test set to measure performance from a predictive point of view. The average results are presented in Table~\ref{tab:perf_out_of_sample}.
\begin{table}[H]
    \centering{%
     \caption{Out-of-sample performance when models are trained with cross-validation (i.e., using both training and validation sets). The adjusted Qini coefficients are averaged over $30$ experiments. Standard-errors are shown in parenthesis. }
    \begin{tabular}{l|ccc}
         \hline
         \bf{Method} & \bf{training-set} & \bf{validation-set} & \bf{test-set} \\
         \hline
         \textit{RF} & $\bf 0.896~(0.031)$ & $0.152~(0.037)$ & $0.071~(0.018)$ \\
         \textit{Q+LHS} & $0.885~(0.032)$ & $\bf 0.859~(0.051)$ & $\bf 0.556~(0.024)$ \\
         \textit{Q+lasso} & $0.618~(0.041)$ & $0.450~(0.030)$ & $0.127~(0.017)$ \\
         \textit{MLE+lasso} & $0.303~(0.037)$ & $0.057~(0.028)$ & $0.049~(0.009)$ \\
         \hline
    \end{tabular}
    \label{tab:perf_out_of_sample}}
\end{table}
Based on these experiments, we see that the \textit{Q+LHS} model gives the best results in terms of prediction.
We are not surprised by the performance of the \textit{RF} model because we had experimented with these \textit{RF} models
in the past, and we have not able to get better predictive performance in other marketing campaign initiatives.

\section{Conclusion}\label{sec:conclusion}

Our goal was to analyze the data of a marketing campaign conducted by an insurance company to retain customers at the end of their contract. A random group of policyholders received an outbound courtesy call made by one of the company's licensed insurance advisors, with the objective to reinforce the customers confidence in the company, to review their coverage and address any questions they might have about their renewal. In the database at our disposal, an independent group of clients was observed and serves as control. In order to evaluate the causal effect of the courtesy call on the renewal or cancellation of the insurance policy of its clients, an uplift model needed to be applied.

We have developed a methodology for estimating parameters of a logistic regression in the context of uplift models.
This is based on a new statistic specially conceived to evaluate uplift models.
The statistic, the adjusted Qini, is based on the Qini coefficient.
It takes into account the correlation between the observed uplift and the predicted uplift by a model.
Maximizing the adjusted Qini to choose an adequate model for uplift acts as a regularizing factor to select parsimonious models, much as lasso does for regression models.

Since the Qini is a difficult statistic to compute, maximizing the adjusted Qini directly is not an easy task. Instead, we proposed to use lasso-type likelihood penalization to search the space of appropriate uplift models, so as to only consider relevant variables for uplift.
Since the usual lasso is not designed for uplift models, we adapted it,
by selecting the value $\hat \lambda$ of the lasso penalization constant that maximizes the adjusted Qini.
At first, this ensures that the selected variables (i.e., those associated with non-zero regression coefficients)
are important variables for estimating uplift.
Then, in a second step, we estimate the parameters that maximize the
adjusted Qini by searching a Latin hypercube sampling (LHS) surface around the lasso estimates.
A variant of this procedure consists of estimating the parameters as those that maximize the likelihood associated with the model selected by $\hat\lambda$, using only the selected variables.

Experimental evaluation showed that for the first stage of the Qini optimized uplift regression,
choosing the penalization constant from the logistic-lasso sequence by maximizing the adjusted Qini
dramatically improves the performance of uplift models. This is the \textit{Q+lasso} model.
In addition, using a LHS search on the second stage leads to a direct maximization of the adjusted Qini coefficient, and to a further boost in the performance of the model.
The resulting model is the \textit{Q+LHS} model.
In addition, our empirical studies clearly show that the performance of a Qini-based regression model is much better than the performance of the usual lasso penalized logistic regression model.

Concerning the particular marketing data available to us from the insurance company, we selected two final models and compared them to the usual lasso regression approach as well as the uplift random forest. The results show that our method clearly surpasses the usual approach in terms of performance. 
We argue that this is due to the Qini-based methods performing variable selection explicitly build for optimizing uplift. Although, even if overall, the marketing campaign of the insurance company did not appear to be successful, the uplift models with the selection of the right variables identify a group of customers for which the campaign worked very well. Indeed, the results show that a persuadable client is a customer with many products to insure. Also, notice there is a subgroup of clients for whom the call had a negative impact. This can be explained by the fact that some customers are already dissatisfied with their insurance policies and have already decided to change them before receiving the call. This call can also trigger a behavior that encourages customers to look for better rates. For future campaigns, the company can target only those customers for whom the courtesy call will be useful and remove and investigate more the clients for whom the marketing campaign had a negative effect.

\section*{Acknowledgements}
Mouloud Belbahri and Alejandro Murua were partially funded by The Natural Sciences and Engineering Research Council of Canada grant~2019-05444. Vahid Partovi Nia
was supported by the Natural Sciences and Engineering Research Council of Canada (NSERC) discovery grant~418034-2012.

\bibliographystyle{plainnat}
\bibliography{mybib}

\end{document}